\documentclass{aastex631}
\usepackage{amsmath}
\usepackage{upgreek}
\usepackage{color}

\newcommand{\pd}{\partial}
\newcommand{\ham}{{\cal H}}
\newcommand{\vect}[1]{\mbox{\boldmath$#1$}}
\newcommand{\grad}{\vect{\nabla}}
\newcommand{\cdiv}[1]{\left( \frac{\pd}{\pd t} +
                            \vect{v} \cdot \grad
                            \right) #1}
\newcommand{\ase}[1]{\tilde{\vect{#1}}}
\newcommand{\hv}[1]{\ensuremath{\hat{\mathbf{#1}}}}
\newcommand{\amp}[1]{\tilde{\cal #1}}
\newcommand{\vpmz}{\varpi_n \!\!\pm \!\!\Delta \omega}
\newcommand{\vpmt}{\varpi_n \!\! \pm \!\! 2\Delta \omega}
\newcommand{\vpn}{\varpi_n}
\newcommand{\vpl}{\varpi_n \!\! + \!\!\Delta \omega}
\newcommand{\vmi}{\varpi_n \!\! - \!\!\Delta \omega}

\accepted{November 21, 2022}

\begin{document}

\title{Polarized Maser Emission with In-Source Faraday Rotation}

\correspondingauthor{T. L. Tobin}
\email{tltobin@umich.edu}

\author[0000-0001-8103-5499]{T. L. Tobin}
\affiliation{Department of Astronomy, University of Michigan, 1085 S. University, Ann Arbor, MI 48109, USA}
\affiliation{Department of Physics, University of Notre Dame,
Nieuwland Science Hall, Notre Dame, IN 46556, USA}
\affiliation{Department of Astronomy, University of Illinois at Urbana-Champaign
1002 W. Green Street, Champaign, IL 61801, USA}

\author[0000-0002-2094-846X]{M. D. Gray}
\affiliation{National Astronomical Research Institute of Thailand
260 Moo 4, T. Donkaew, A. Maerim, Chiangmai 50180, Thailand}
\affiliation{Jodrell Bank Centre for Astrophysics, Department of Physics and Astronomy
University of Manchester, M13 9PL, UK}

\author[0000-0001-6233-8347]{A. J. Kemball}
\affiliation{Department of Astronomy, University of Illinois at Urbana-Champaign
1002 W. Green Street, Champaign, IL 61801, USA}

\begin{abstract}
We discuss studies of polarization in astrophysical masers with particular emphasis on
the case where the Zeeman splitting is small compared to the Doppler profile, resulting in
a blend of the transitions between magnetic substates. A semi-classical theory of the 
molecular response is derived, and coupled to radiative transfer solutions for 1 and 2-beam
linear masers, resulting in a set of non-linear, algebraic equations for elements of the molecular
density matrix. The new code, PRISM, implements numerical methods to compute these solutions. Using PRISM, we demonstrate a smooth transfer
between this case and that of wider splitting. For a J=1-0 system, with parameters based
on the $v=1, J=1-0$ transition of SiO, we investigate the behaviour of linear and circular
polarization as a function of the angle between the propagation axis and the magnetic field, and
with the optical depth, or saturation state, of the model. We demonstrate how solutions are
modified by the presence of Faraday rotation, generated by various abundances of free electrons,
and that strong Faraday rotation leads to additional angles where Stokes-Q changes sign.
We compare our results to a number of previous models, from the analytical limits derived
by Goldreich, Keeley and Kwan in 1973, through computational results by W. Watson and co-authors,
to the recent work by Lankhaar and Vlemmings in 2019. We find that our results are generally
consistent with those of other authors given the differences of approach and the approximations
made.

\end{abstract}


\section{Introduction} 
\label{s:intro}

Polarization in astrophysical masers has been one of the most controversial themes in
the field, owing to a number of rival theories and further obfuscation due to different
conventions regarding the definitions of right and left-handed polarized radiation, and the Stokes
parameters that are widely used to describe intensity-like quantities. The early paper
by \citet{1973ApJ...179..111G} (GKK) is still regarded as the seminal work in maser polarization theory,
at least where this is based on Zeeman splitting of molecular transitions. It separates possible
masers into a number of cases, based on the strength of the magnetic field and the degree
of saturation, for example. Perhaps the most important limit, with regard to the present work, is
the case where the magnetic field is strong enough to define a good quantization axis, but is adequate 
only to split the transition by a frequency much smaller than the Doppler width. It is in this case,
where the Zeeman-split transitions form an overlapping group, that most controversy has arisen. GKK
analysed this case in the limit of ultimate saturation, where the differentials of the Stokes
parameters in the maser propagation equations tend to zero, and were able to obtain a set of analytical
expressions for the Stokes parameters.

What GKK did not do was to analyse the overlapping group in the intermediate saturation regime.
To do this, the differential equations describing maser amplification and saturation must be solved
in a consistent manner all the way from negligible saturation to very high degrees of saturation. Numerical solutions covering the required range of saturation were
computed in a series of papers involving the late W.D. Watson,
beginning with \citet{1984ApJ...285..158W}, which included models with two
counter-propagating beams and both $J=1-0$ and $J=2-1$ Zeeman systems. 
An important result was
that the GKK limits for strong saturation were approached rather slowly, and
are the same for $J=2-1$ and $J=1-0$ groups. The model was later extended to
include the effects of velocity gradients \citep{1986ApJ...302..108D} and applied
specifically to the 22-GHz maser transition of H$_2$O \citep{1986ApJ...302..750D}.
Further investigations found that a high polarization regime, 
originally investigated by GKK, where the stimulated
emission rate, R, exceeds the Zeeman splitting, $g\Omega$, but is vastly less
than the square of the splitting divided by the loss rate, $\Gamma$ (i.e. $g\Omega \ll R \ll (g\Omega)^2/\Gamma$), applies only to
the $J=1-0$ system. The usual restriction on the stimulated emission rate to
be smaller than the Zeeman splitting was lifted \citep{1990ApJ...354..660N} by
including off-diagonal elements of the density matrix that couple Zeeman
substates of levels with the same value of $J$: we will refer to these later as type~2 elements.

The early Watson models had boxcar line profiles and no spectral information, and
could therefore not address the generation of circular polarization in
the overlapping Zeeman case: Stokes $V$ is zero at line center, and antisymmetric
about the center, so that a value of zero also results from a line profile
average.
Circular polarization and a spectral response were added in improved models
that computed Stokes $V$ \citep{1992ApJ...384..185N,1994ApJ...423..394N}.
Anisotropic pumping, a feature of many numerical models since 
\citet{1984ApJ...285..158W} was favoured over Faraday depolarization for 
selective loss of Stokes $Q$ and $U$ \citep{1997ApJ...481..832W}. Maser
polarization from unsaturated amplification through a turbulent velocity field
driven by the rotation of an accretion disc was studied in 
\citet{2001ApJ...557..967W}. The work from this series on which we base most
of our comparisons is \citet{2001ApJ...558L..55W}, which is based on the
\citet{1992ApJ...384..185N} model, but includes calculations at many more angles
of the magnetic field with respect to the radiation propagation axis, and a wider
range of saturation levels. We note here that although the fundamental equations
that we use are the same as those used in the Watson series, there are
substantial differences in the methods of solution, so a primary purpose of
the current work is to demonstrate that very similar results arise from these
different methods. One important difference is that models in the Watson series
are solved via a time-domain molecular polarization
(for example eq.(11) of \citet{1990ApJ...354..660N}), followed presumably by a
steady-state approximation to their eq.(4), whilst our method involves a formal Fourier
transform to the frequency domain, where the combined density matrix and
radiative transfer equations are solved: a spectral distribution is therefore
fundamental to our model. Our method therefore has more in common with the
methods used by \citet{1978PhRvA..17..701M} (no polarization) 
and \citet{2009MNRAS.399.1495D} than with Watson and
his co-workers. A second important
difference is in the way in which the radiation transfer itself is treated:
while the calculations in the Watson series are based on the solution of
coupled ODEs, we make formal solutions of the transfer equations, reducing the
problem to a set of non-linear algebraic equations in the inversions
(see Section~\ref{s:radtran}).

Another problem that bedevils polarisation work in general is that authors, over
the years, have adopted several different conventions regarding definitions of
left and right-handed waves, the definition of Stokes $V$ and the labeling of the
helical transitions within the Zeeman pattern. Although most work is internally
self-consistent, it is often confusing to relate it to the theoretical work of
others, and to observational data. These problems are discussed in
\citet{2014MNRAS.440.2988G}, where maser polarisation 
conventions are discussed in relation to
observations of polarization in the 21-cm hydrogen line. We specify our
conventions in Section~\ref{s:satn}.

\subsection{Application to Observations}
With regard to observational data,
strong polarization is one of the characteristics of astrophysical masers, detected during the
earliest work in the field \citep{1965Natur.208..440W}. 
A good summary is \citet{2018IAUS..336...27S}. A distinction should be drawn
between the paramagnetic molecules, for example OH, CH, in which much information can be gleaned directly
from observations, and the closed shell species (for example H$_2$O, SiO and CH$_3$OH) in which more
sophisticated analysis, including numerical modeling, is generally needed. In the former case, the
magnitude of the magnetic field can typically be determined from the Zeeman splitting of lines, rather
than their polarization, and the sense of the magnetic field (towards or away from the observer) can
be determined from the handedness of elliptical polarization found in the lower-frequency member of
a pair, for example \citet{2014MNRAS.440.2988G}. If linear polarization is present, the 
orientation of the magnetic field in the plane of the sky can be deduced, providing that a spectral component can be identified as a sigma or
pi transition. Radiative transfer analysis is only required for a full 3D reconstruction of the vector
magnetic field. 

In the case of closed-shell molecules, the Zeeman splitting is much smaller than the
Doppler line width, and numerical modeling is required to extract any useful information at all from
polarization-sensitive observations. For example, direct observables, like the Stokes-I and
Stokes-V spectral profiles need model fitting to recover the line-of-sight component of the
magnetic field and further modeling to recover the full field strength, since the relation
that the fractional circular polarization is proportional to $\cos \theta$ breaks down for saturated masers 
(for example \citealt{2006A&A...448..597V}). The
derived angle $\theta$, between the magnetic field and the maser propagation direction, and the fractional linear polarization, may then be used to derive the level of
saturation of the maser. Knowledge of $\theta$ can also be used to break the EVPA (electric vector
position angle) degeneracy, determining
the field as either parallel or perpendicular to the EVPA \citep{2006A&A...448..597V}.
Small-scale (of order tens to hundreds of AU) variations in field structure can be traced if the field direction can be followed along
imaged maser features. Examples of this include preferential alignment of the magnetic field with outflow axes in
massive star-forming regions \citep{2015A&A...578A.102S}, and the change in field orientation over
a timescale of 7\,yr in the VLA2 sub-source of W75N \citep{2014A&A...565L...8S}.

Observations of SiO masers towards asymptotic giant branch (AGB) stars have revealed cases where
the EVPA of linear polarization rotates through approximately $\pi/2$ within the apparent confines
of a single maser object or cloud \citep{2011ApJ...743...69K,2013MNRAS.431.1077A}. Modeling
of such EVPA rotations offers the possibility of distinguishing between the Zeeman interpretation
of maser polarization and a number of competing theories \citep{2019ApJ...871..189T}. The phenomenon
may also be related to pulsation shocks emanating from the star and/or to the overall
magnetic field structure of the circumstellar envelope that can be tested through additional
models, for example \citet{2010PASP..122.1334P,2020PASP..132c4203P}.

\section{Saturation model}
\label{s:satn}

Our model is derived through the following key stages: First, the time-dependent 
Schr\"{o}dinger equation is solved via an expansion of the wavefunction in a basis set of the
eigenfunctions (the energy levels) of the corresponding time-independent equation. Products of the
coefficients of this expansion, averaged over
a volume of order $\lambda^3$, where $\lambda$ is a typical maser wavelength, become elements of
the density matrix (DM). This solution is facilitated by a separation of the Hamiltonian operator
into a time-independent component and a time-dependent interaction component that is a function of the
electric field that drives transitions between the energy levels. If this interaction operator is
further split into a coherence-preserving component, based on the electric field of the maser, and another component containing all other (`kinetic') processes, then the solution of the original 
Schr\"{o}dinger equation reduces to solving a pair of differential equations for elements of the DM:
one for diagonal elements, where the energy-level indices of the element are equal, and one
for off-diagonal elements, where they are not. With a little more work, diagonal elements may be
paired, resulting in equations for the inversion between pairs of levels. The resulting
`optical Bloch equations' may be written,
\begin{equation}
    \cdiv \rho_{pq} = \frac{i}{\hbar} \sum_{j=1,\neq q,p}^N \left(
          \rho_{pj} \ham_{qj}^* - \rho_{jq} \ham_{pj}
                                                            \right)
                                                            +
    \frac{i\ham_{pq}}{\hbar} \Delta_{pq} - (\gamma_{pq} + i\omega_{pq} )\rho_{pq},
\label{bloch_off}    
\end{equation}
for the off-diagonal DM element, $\rho_{pq}$, representing coherence between levels $p$ and $q$, and
\begin{equation}
   \cdiv \Delta_{pq} = -\frac{2}{\hbar} \Im \left\{
      2\rho_{pq} \ham_{qp} + \sum_{j=1,\neq p,q}^N \left(
         \rho_{pj} \ham_{jp} - \rho_{qj} \ham_{jq}
                                                   \right)
                                            \right\}
                    - \Gamma_{pq} \Delta_{pq} + P_{pq} \phi (\vect{v}),
\label{bloch_diag}   
\end{equation}
for the population inversion, $\Delta_{pq}$, between these levels. All DM elements are functions of time,
$t$, position, $\vect{r}$ and Doppler velocity, $\vect{v}$. The index $j$ runs over the $N$ energy levels in the model, and
angular frequencies of the transitions between them are written $\omega_{pq}=(E_p - E_q)/\hbar$. 
The maser part of the interaction hamiltonian, linking levels $p$ and $q$, is represented as
$\ham_{pq}$. As matrices, both the interaction hamiltonian and the DM are hermitian. The symbol
$\Im$ denotes taking the imaginary part. Coherence between
levels is lost at the rate $\gamma_{pq}$, which encompasses all elastic and inelastic collisions, and
radiative processes that are not stimulated emission across the maser levels. $\Gamma_{pq}$ is the loss rate
to the inversion, and so represents a subset of those processes contributing to $\gamma_{pq}$, since
elastic processes are excluded. A phenomenological pump rate per unit volume, $P_{pq} \phi(\vect{v})$, is included to
support the inversion. The pumping term contains the normalised gaussian
function $\phi(\vect{v})$ with width parameter,
\begin{equation}
w = \sqrt{2 k_B T_k / m_X},
\label{eq:vwidth}
\end{equation}
where $k_B$ is Boltzmann's constant, $T_k$ is the kinetic temperature in the
maser zone and $m_X$ is the molecular mass of the maser species. We assume negligible velocity
redistribution (NVR), so that population is not exchanged between different velocity subgroups of the DM.
This approximation also implies that $\gamma_{pq} \sim \Gamma_{pq}$.

The maser part of the interaction hamiltonian is defined as
\begin{equation}
\ham_{pq} = - \hat{\vect{d}}_{pq} \cdot \vect{E},
\label{hamdef}
\end{equation}
where $\hat{\vect{d}}_{qp},$ is the molecular dipole operator for the $pq$ transition, and $\vect{E}$ 
is the electric field of the maser radiation. In the case of Zeeman-based maser polarization, we do
not assign the electric field to a particular transition, since the response of several transitions
may overlap in frequency. The electric field $\vect{E}$ is the real part of the complex analytic 
signal $\ase{E}$. From this point, we consider propagation only along the $z$-axis, so that the analytic
signal appears as 
\begin{equation}
\ase{E} (z,t) = ( \amp{E}_x \hv{x} + \amp{E}_y \hv{y}) e^{-i\omega (t-z/c)} ,
\label{asig1}
\end{equation}
for propagation in the positive $\hv{z}$ direction, where $\amp{E}_x$ and $\amp{E}_y$ are the $x$ and $y$ components, respectively, of 
the time-domain complex amplitude of
the field. We consider a broad-band electric field, and this
may be written decomposed into its Fourier components of width $2\pi /T$, where $T$ is a finite sampling
time. In the Fourier representation,
\begin{equation}
\ase{E} (z,t) = \sum_{n=1}^\infty (\amp{E}_{x,n} \hv{x} + \amp{E}_{y,n} \hv{y}) e^{-i\omega_n (t-z/c)} ,
\label{asig2}
\end{equation}
where $\omega_n$ is the angular frequency of the $n$th Fourier component and $\amp{E}_{x,n}$ and $\amp{E}_{y,n}$ are the components of the field amplitude at that frequency.

Our conventions regarding the electric field are the following: we adopt the IEEE definition of a right-handed
wave (IEEE-STD-145 of 1993) and the IAU axis system, in which the $z$ axis points towards
the observer, the $x$ axis towards North and the $y$ axis, East
\citep{1996A&AS..117..161H,2010PASA...27..104V}. We also use the IAU definition of Stokes $V$ as the 
right-handed intensity minus the left-handed intensity. We use the naming
convention of \cite{1988ApJ...326..954G} for Zeeman-split transitions, 
so that, for a molecule
that has Land\'{e} factors with the same sign as OH, the $\sigma^+$ transition
has the lowest frequency in a triplet,
and changes magnetic quantum number by $+1$ in
emission.

The second key stage is the application of a rotating wave approximation that eliminates all terms oscillating
at frequencies corresponding to the band-centre frequency of the radiation. To this end, the Fourier frequencies
in eq.(\ref{asig2}) are expanded as $\omega_n = \omega_0 + \varpi_n$, where $\omega_0$ is the band-center frequency
and $\varpi_n$ is a local frequency, measured from $\omega_0$, and is never larger than a few Doppler widths.

The third stage is to apply a time-to-(angular) frequency Fourier transform of the time-domain DM equations
\citep{1978PhRvA..17..701M,2009MNRAS.399.1495D}. The result is a set of non-linear algebraic equations in the
Fourier components of the inversions and of the slow part of the off-diagonal DM equations. These equations also
contain Fourier components of the electric field complex amplitudes. 

At this point, we restrict our analysis to the case where there is just one transition of each helical
type, which requires both unsplit maser levels to have the 
same Land\'{e} splitting factor. We do this because the numerical calculations in the present work apply to the Zeeman-split
$J=1-0$ transition of SiO. More general cases have Zeeman energy shifts that are different in the upper
and lower unsplit states, for example eq.(9.1) of \citet{mybook}. In this case there can be several
transitions of each helical type when the external magnetic field is applied, and these have different
line strengths in general. The formalism of a single transition each of $\pi$, $\sigma^+$ and $\sigma^-$
type can be restored by employing the averaged Land\'{e} splitting factor for one group of $\sigma$
transitions (by symmetry the average for the $\pi$ transitions is zero). See, for example
\citet{2004ASSL..307.....L}.

Dipole operators
that are pure right- or left-handed for the respective $\sigma^+$ and
$\sigma^-$ Zeeman 
transitions, or lie along the $z'$-axis for $\pi$ transitions in a frame where the
magnetic field is $\vect{B}=B\hv{z}'$, have been rotated into a frame where the
radiation propagates along the $z$ axis. To allow for Faraday rotation, the
$x'$ axis is constrained to lie in the $xy$ plane, but may be offset by an
angle $\phi$ from the $x$ axis. The angle $\theta$ is the
offset between the $z'$ and $z$ axes. 

The slow off-diagonal DM elements are
then either
\begin{equation}
s_n^\pm = \frac{\pi \hat{d}^\pm \tilde{L}^\pm}{\hbar} \sum_{m=-\infty}^\infty \Delta_{n-m}^\pm
        \left\{
         \amp{E}_{R,m} (1 \mp \cos \theta ) (\cos \phi + i \sin \phi)
       + \amp{E}_{L,m} (1 \pm \cos \theta ) (\cos \phi - i \sin \phi )
        \right\},
\label{eq:dmoff1}
\end{equation}
for $\sigma$ transitions, where superscripts $\pm$ denote the transition type, and
subscripts $m,n$, or combinations thereof, label the Fourier components. The
electric field amplitudes in eq.(\ref{eq:dmoff1}) are in the circular polarization
basis, with $R,L$ denoting right and left-handed polarization, respectively.
The symbol $\tilde{L}^\pm$ denotes a complex lorentzian function. For $\pi$
transitions, identified by the zero superscript,
\begin{equation}
 s_n^0 = \frac{\sqrt{2} \pi \hat{d}^0 \tilde{L}^0}{\hbar}
          \sum_{m=-\infty}^\infty \Delta_{n-m}^0
           \left\{
              \amp{E}_{R,m} (\sin \phi - i \cos \phi)
             +\amp{E}_{L,m} (\sin \phi + i \cos \phi) 
           \right\}\sin \theta .
         \label{eq:dmoffpi1}
\end{equation}
The inversions,
$\Delta^{\pm}_n$ and $\Delta^0_n$ are now also 
labeled by Fourier component, or population
pulsation, $n$, in the $\sigma^{\pm}$ and $\pi$ transitions.
We note that the inversion in the central Fourier component, $n=0$ is real, but
the other pulsations are complex in general. Fourier components of the inversion
are given by the diagonal DM equations,
\begin{align}
\Delta_n^\pm & = 2\pi \amp{L}^\pm P^\pm \phi(v) \delta_n \nonumber \\ &-\frac{\pi \amp{L}^\pm}{2\hbar}\sum_{m=-\infty}^\infty
        \left\{ \frac{\sin \theta}{\sqrt{2}}
            \left[
              \left(  \hat{d}^{0*} s_{m+n}^0  \amp{E}_{R,m}^*
                     + \hat{d}^0 s_{m-n}^{0*} \amp{E}_{L,m}
              \right) (s+ i c)
             +\left( \hat{d}^{0*} s_{m+n}^0  \amp{E}_{L,m}^* 
                     + \hat{d}^0 s_{m-n}^{0*} \amp{E}_{R,m}
              \right) (s - i c)
            \right]
           \right. \nonumber \\
          & \left. +\frac{(1\pm \cos \theta)}{2}
             \left[
              \left( 2 \hat{d}^{\pm *} s_{m+n}^\pm \amp{E}_{L,m}^*
                      +\hat{d}^\mp s_{m-n}^{\mp *} \amp{E}_{R,m}
              \right) (c+i s)
             +\left( 2 \hat{d}^\pm s_{m-n}^{\pm *} \amp{E}_{L,m}
                      +\hat{d}^{\mp *} s_{m+n}^\mp \amp{E}_{R,m}^*
              \right) (c-i s)
            \right]                      
           \right. \nonumber \\
          & \left. +\frac{(1\mp \cos \theta)}{2} 
            \left[
          \left( 2 \hat{d}^{\pm *} s_{m+n}^\pm \amp{E}_{R,m}^*
                  +\hat{d}^\mp s_{m-n}^{\mp *} \amp{E}_{L,m}
          \right) (c-i s)
          +\left( 2 \hat{d}^\pm s_{m-n}^{\pm *}\amp{E}_{R,m}
                   +\hat{d}^{\mp *} s_{m+n}^\mp \amp{E}_{L,m}^*
            \right) (c+i s)
             \right]                                  
        \right\}
\label{eq:diag1sig}
\end{align}
for the $\sigma^\pm$ transitions, where $c+is = \cos \phi + i \sin \phi$, $\delta_n$ is the Kronecker delta and
\begin{align}
\Delta_n^0 & = 2\pi \amp{L}^0 P^0 \phi(v) \delta_n \nonumber \\ 
&-\frac{\pi \amp{L}^0}{2\hbar}\sum_{m=-\infty}^\infty
        \left\{
           \sqrt{2} \sin \theta 
             \left[
               \left( \hat{d}^{0*} s_{m+n}^0 \amp{E}_{R,m}^*
                     +\hat{d}^0 s_{m-n}^{0*} \amp{E}_{L,m}
               \right) (s+i c)
              +\left( \hat{d}^{0*} s_{m+n}^0 \amp{E}_{L,m}^* 
              +       \hat{d}^0 s_{m-n}^{0*} \amp{E}_{R,m}
               \right) (s-i c)
            \right]
           \right. 
    \nonumber \\
          & \left. +\frac{(1 + \cos \theta)}{2} 
             \left[ 
                \left( \hat{d}^{+ *} s_{m+n}^+ \amp{E}_{L,m}^*
                      +\hat{d}^- s_{m-n}^{-*} \amp{E}_{R,m}
                \right) (c+i s)
               +\left( \hat{d}^+ s_{m-n}^{+*} \amp{E}_{L,m}
                       +\hat{d}^{- *} s_{m+n}^- \amp{E}_{R,m}^*
                \right) (c-i s)
             \right]    
            \right. 
    \nonumber \\
          & \left. +\frac{(1 - \cos \theta)}{2} 
             \left[
               \left( \hat{d}^{+ *} s_{m+n}^+ \amp{E}_{R,m}^* 
                     +\hat{d}^- s_{m-n}^{-*} \amp{E}_{L,m}
               \right) (c-i s)
              +\left( \hat{d}^+ s_{m-n}^{+*} \amp{E}_{R,m}
                    + \hat{d}^{- *} s_{m+n}^- \amp{E}_{L,m}^*
               \right) (c+i s)
            \right]
        \right\}
\label{eq:diag1pi}
\end{align}
for $\pi$ transitions. 

Further operations carried out on 
eq.(\ref{eq:diag1sig}) and eq.(\ref{eq:diag1pi}), and
the definitions of the complex lorentzian functions,
$\amp{L}^0$ and $\amp{\cal L}^\pm$ are deferred to
Appendix~\ref{a:appendix1}. The results of these operations are the
key equations for the molecular response as a function of angular frequency,
\begin{align}
\rho_{n\pm k}^\pm & = \frac{P^\pm \phi_{n\pm k}}{\Gamma^\pm}
     - \frac{\pi}{4\hbar^2 \Gamma^\pm c \epsilon_0}
   \left\{
       2|\hat{d}^0|^2 \rho_{n\pm k}^0
          \left[
            I_{n\pm k} - Q_{n\pm k} \cos(2\phi) - U_{n\pm k}\sin(2\phi)
          \right] \sin^2 \theta 
   \right. \nonumber \\
 & \left. +
      |\hat{d}^\mp|^2 \rho_{n\pm k}^\mp 
          \left[
             (1+\cos^2 \theta) I_{n\pm 2k} 
              + Q_{n\pm 2k}\sin^2 \theta \cos (2\phi)
              + U_{n\pm 2k}\sin^2 \theta \sin (2\phi)
              \pm 2 V_{n\pm 2k} \cos \theta
          \right]
   \right. \nonumber \\
 & \left. +
   2 |\hat{d}^\pm|^2 \rho_{n\pm k}^\pm
          \left[
            (1+\cos^2 \theta) I_n
            + Q_n \sin^2 \theta \cos (2\phi)
            + U_n \sin^2 \theta \sin(2\phi)
            \mp 2 V_n \cos \theta
          \right]
   \right\}
\label{eq:sig_key}
\end{align}
for the $\sigma^\pm$ transitions and
\begin{align}
\rho_{n}^0 & = \frac{P^0\phi_n}{\Gamma^0}
     - \frac{\pi}{4 \hbar^2 \Gamma^0c \epsilon_0}
     \left\{
     4|\hat{d}^0|^2 \rho_n^0
          \left[
            I_n - Q_n \cos(2\phi) - U_n \sin(2\phi)
          \right] \sin^2 \theta
     \right. \nonumber \\
  & \left. +
      |\hat{d}^-|^2 \rho_n^-
          \left[
             (1+\cos^2 \theta) I_{n+ k} 
              + Q_{n+k}\sin^2 \theta \cos (2\phi)
              + U_{n+k}\sin^2 \theta \sin (2\phi)
              + 2 V_{n+k} \cos \theta
          \right]
     \right. \nonumber \\
  & \left. +
      |\hat{d}^+|^2 \rho_n^+
           \left[
              (1+\cos^2 \theta) I_{n-k} 
              + Q_{n-k}\sin^2 \theta \cos (2\phi)
              + U_{n-k}\sin^2 \theta \sin (2\phi)
              - 2 V_{n-k} \cos \theta
           \right]
     \right\} ,
\label{eq:pi_key}
\end{align}
for the $\pi$ transition,
noting that molecular responses are now saturated by Stokes parameters
at specified frequencies. The index $k$ denotes a shift in frequency bins corresponding to the Zeeman splitting. We derive expressions in the following 
section that allow us to eliminate the Stokes parameters from
eq.(\ref{eq:sig_key}) and eq.(\ref{eq:pi_key}).

\section{Radiative transfer solution}
\label{s:radtran}

At this point, our treatment departs significantly from most earlier
work on maser polarization: instead of solving a set of differential
radiative transfer equations in the Stokes parameters, we generate
instead a formal solution of the transfer equation and eliminate the
Stokes parameters, to leave only a set of algebraic equations in the
elements of the DM.

In scalar radiative transfer problems, the elimination of the
radiation intensity is one of the oldest methods of solving the
combined radiative transfer and non-LTE statistical balance problem, for example \citet{1950ratr.book.....C,1964ApJ...139..397K},
and is useful because it reduces the problem to a set of non-linear
algebraic equations in the molecular populations or inversions
only, obviating the need to compute any radiation integrals. The
method was introduced in a modern context for the classic slab
geometry by \citet{2006MNRAS.365..779E}, and has been used successfully
for masers in a 3D finite-element model of maser clouds
\citep{2018MNRAS.477.2628G,2019MNRAS.486.4216G,2020MNRAS.493.2472G}.

The additional problem, in the context of the present work, is that
the standard method of writing a formal solution of the radiative
transfer equation via the integrating factor method is not
obviously applicable to the vector-matrix radiative transfer
equation,
\begin{equation}
d\vect{I}_n / ds = \upgamma_n \vect{I}_n
\label{eq:rteq}
\end{equation}
required to propagate polarized radiation along ray $s$.

To obtain a formal solution of eq.(\ref{eq:rteq}) with the Stokes
vector, $\vect{I}_n$, as the subject, we have 
adopted the operator method
proposed by \citet{1987nrt..book..265L}. Following this method, the
solution of eq.(\ref{eq:rteq}) is
\begin{equation}
\vect{I}_n (s) = \mathrm{O}_n (s,s') \vect{I}_n (s'),
\label{eq:opsoln}
\end{equation}
for moving from ray position $s'$ to position $s$, where the
evolution operator is the $4\times 4$ matrix,
\begin{equation}
\mathrm{O}_n (s,s') = \mathrm{U} + 
   \sum_{M=1}^\infty \frac{1}{M!} \int_{s'}^{s} ds_1
                                  \int_{s'}^{s} ds_2
                                  \dots
                                  \int_{s'}^{s} ds_M
    P \left\{
         \upgamma_n(s_1) \upgamma_n(s_2) ... \upgamma_n(s_M)
      \right\},
\label{eq:evop}
\end{equation}
where $\mathrm{U}$ is the identity matrix. Note that the operator
in eq.(\ref{eq:evop}) differs from a standard exponential because
of the presence of $P$, the `chronological' operator, that specifies
the order in which the various gain matrices, 
the $\upgamma_n(s_M)$, must be
multiplied.

The gain matrices used in the present work are the sum of a continuum
version, containing only the Faraday rotation elements, and a line
version, containing all the others. The combined gain matrix is
\begin{align}
\upgamma_n(s) = \left[
   \begin{array}{cccc}
      \gamma_{I,n}   &  -\gamma_{Q,n}  &  -\gamma_{U,n}  &  -\gamma_{V,n} \\
     -\gamma_{Q,n}   &   \gamma_{I,n}  &   \gamma_{QU,n} &      0         \\
     -\gamma_{U,n}   &  -\gamma_{QU,n} &   \gamma_{I,n}  &      0         \\
     -\gamma_{V,n}   &        0        &        0        &   \gamma_{I,n}
   \end{array}
                \right],
\label{eq:gainmat}
\end{align}
where the individual elements are defined as
\citep{1976A&AS...25..379L,1982SoPh...78..355L,1987nrt..book..213R}
\begin{align}
\gamma_{I,n} & = \frac{\pi}{4\epsilon_0 \hbar} \left[
   2 |\hat{d}^0|^2 \rho_n^0 \sin^2 \theta 
 +   |\hat{d}^+|^2 \rho_{n+k}^+ (1 + \cos^2 \theta)
 +   |\hat{d}^-|^2 \rho_{n-k}^- (1 + \cos^2 \theta)
                                             \right]
\label{eq:gammai} \\
\gamma_{Q,n} & = \frac{\pi}{4\epsilon_0 \hbar} \left[
    2 |\hat{d}^0|^2 \rho_n^0
  -   |\hat{d}^+|^2 \rho_{n+k}^+
  -   |\hat{d}^-|^2 \rho_{n-k}^-
                                             \right] 
                                             \sin^2 \theta \cos(2\phi)
\label{eq:gammaq} \\
\gamma_{U,n} & = \frac{\pi}{4\epsilon_0 \hbar} \left[
    2 |\hat{d}^0|^2 \rho_n^0
  -   |\hat{d}^+|^2 \rho_{n+k}^+
  -   |\hat{d}^-|^2 \rho_{n-k}^-
                                             \right] 
                                             \sin^2 \theta \sin(2\phi)
\label{eq:gammau} \\
\gamma_{V,n} & = \frac{2\pi}{4\epsilon_0 \hbar} \left[
      |\hat{d}^+|^2 \rho_{n+k}^+
  -   |\hat{d}^-|^2 \rho_{n-k}^-
                                             \right]
                                             \cos \theta 
\label{eq:gammav} \\
\gamma_{QU,n} & = -\frac{e^3 n_e B \cos \theta}
                        {8\pi^2 \epsilon_0 m_e^2 c \nu_c^2}.
\label{eq:gammaqu}
\end{align}
In the Faraday term, eq. (\ref{eq:gammaqu}), $n_e$ is the number density of free
electrons, $m_e$ is the electron rest mass and $\nu_c$ is the band-center
frequency of the maser radiation. We neglect continuum processes that would convert between linear and circular polarization.

We now note that if the evolution operator is eliminated 
from eq.(\ref{eq:opsoln}), with the aid of eq.(\ref{eq:evop}), a formal
solution of the form,
\begin{equation}
\vect{I}_n (s) = \vect{I}_n(0) + \int_0^s ds_1 \upgamma_n(s_1) \vect{I}_n(0)
  + \frac{1}{2!} \int_0^s ds_1 \int_0^s ds_2 \upgamma_n(s_1) \upgamma_n(s_2)
  \vect{I}_n(0) + \dots
\label{eq:fullformsol}
\end{equation}
is obtained, where the right-hand side is independent of the Stokes parameters
except for background values where radiation enters the maser zone. For the 
case of an unpolarized background, $\vect{I}_n(0) = (I_{BG},0,0,0)^T$, where
$I_{BG}$ is a background specific intensity.
Furthermore, from the expressions in eq.(\ref{eq:gammai})-eq.(\ref{eq:gammav}),
we see that elimination of the Stokes parameters from 
eq.(\ref{eq:sig_key}) and eq.(\ref{eq:pi_key}) leads to a 
set of integral equations in the molecular
responses at a particular position in the maser column. When the integrals
are replaced by finite sums for numerical work, these equations become
sets of non-linear algebraic equations involving molecular responses at,
in principal, all positions, but no variable radiation intensities.

In order to make the non-linear algebraic equations suitable for numerical
solution, eq.(\ref{eq:sig_key}) and eq.(\ref{eq:pi_key}) were reduced to
the respective dimensionless forms,
\begin{align}
\delta_{n\pm k}^\pm = \alpha^\pm e^{-\varpi^2_{n\pm k}/W^2} & - \left\{
    2\delta_{n\pm k}^0 \left[
       i_{n\pm k} - q_{n\pm k} \cos 2\phi - u_{n\pm k} \sin 2\phi
                      \right] \sin^2 \theta                    \right.
                      \nonumber \\
  &                                                           \left. +
    \eta^\mp \delta_{n\pm k}^\mp \left[
       (1+\cos^2 \theta) i_{n\pm 2k} 
       +(q_{n\pm 2k}\cos 2\phi + u_{n\pm 2k}\sin 2\phi)\sin^2 \theta
       \pm 2v_{n\pm 2k}\cos \theta   
                      \right]                                  \right.
                      \nonumber \\
  &                                                            \left. +
    2\eta^\pm \delta_{n\pm k}^\pm \left[
       (1+\cos^2 \theta) i_n
       +(q_n \cos 2\phi + u_n \sin 2\phi)\sin^2 \theta
       \mp 2v_n \cos \theta
                                  \right]
                                                               \right\}
\label{eq:sig_dimless}
\end{align}
and
\begin{align}
\delta_n^0 = e^{-\varpi^2_n/W^2} & - \left\{
    4\delta_n^0 \left[
       i_n - q_n \cos 2\phi -u_n \sin 2\phi
                 \right] \sin^2 \theta \right.
            \nonumber \\
   &                                   \left. +
    \eta^- \delta_n^- \left[
        (1+\cos^2 \theta) i_{n+k}
        +(q_{n+k}\cos 2\phi + u_{n+k}\sin 2\phi)\sin^2 \theta
        +2v_{n+k}\cos \theta
                      \right]        \right.
            \nonumber \\
   &                                 \left.  +
    \eta^+ \delta_n^+ \left[
         (1+\cos^2 \theta) i_{n-k}
         +(q_{n-k}\cos 2\phi + u_{n-k}\sin 2\phi)\sin^2 \theta
         -2v_{n-k}\cos \theta
                       \right]
                                     \right\},
\label{eq:pi_dimless}
\end{align}
where $W$ is the Doppler width ($w$ in velocity units) converted to angular frequency. The lower case Stokes parameters $i,q,u,v$ are scaled to the saturation
intensity, assumed here to be the same for all three helical transitions, and
equal to,
\begin{equation}
I_{sat} = \frac{16 h c \Gamma}
               {3\lambda_0^3 A^0},
\label{eq:isat}    
\end{equation}
where $\Gamma$ is the loss rate and $\lambda_0$ and $A^0$ are the rest
wavelength and Einstein A-value of the $\pi$-transition.
The dimensionless Stokes
parameters are eliminated from eq.(\ref{eq:sig_key}) and eq.(\ref{eq:pi_key})
with the help of scaled versions of eq.(\ref{eq:fullformsol}) in which
distances are scaled to an optical depth $\tau$, defined as
\begin{equation}
d\tau = \frac{h c P^0}
             {2 \pi^{3/2} w I_{sat}} ds .
\label{eq:dtau}
\end{equation}
Molecular responses have been scaled according to
\begin{equation}
\delta_n^{\pm ,0} = \frac{\sqrt{\pi} w \Gamma}
                         {P^0} \rho_n^{\pm ,0},
\label{eq:deltadef}
\end{equation}
in the process of converting eq.(\ref{eq:sig_key}) (eq.(\ref{eq:pi_key})) to
eq.(\ref{eq:sig_dimless}) (eq.(\ref{eq:pi_dimless})). Other dimensionless
parameters appearing in eq.(\ref{eq:sig_dimless}) and eq.(\ref{eq:pi_dimless})
are $\alpha^\pm = P^\pm /P^0$ the relative pump rates in $\sigma$ 
and $\pi$ transitions, and $\eta^\pm = |\hat{d}^\pm|^2 / |\hat{d}^0|^2$, the
corresponding relative line strengths. When applied to
eq.(\ref{eq:gammai})-eq.(\ref{eq:gammav}) the scalings 
in eq.(\ref{eq:isat})-eq(\ref{eq:deltadef})
result in the gain matrix elements
\begin{align}
\gamma_{i,n} & = 2\delta_n^0 \sin^2 \theta 
            + (\eta^+ \delta_{n+k}^+ +\eta^- \delta_{n-k}^-)(1+\cos^2 \theta)
            \label{eq:gi_dimless}\\
\gamma_{q,n} & = (2\delta_n^0 - \eta^+ \delta_{n+k}^+ - \eta^- \delta_{n-k}^-)
                \sin^2 \theta \cos 2\phi
            \label{eq:gq_dimless}\\
\gamma_{u,n} & = (2\delta_n^0 - \eta^+ \delta_{n+k}^+ - \eta^- \delta_{n-k}^-)
                \sin^2 \theta \sin 2\phi
            \label{eq:gu_dimless}\\
\gamma_{v,n} & = 2 (\eta^+ \delta_{n+k}^+ - \eta^- \delta_{n-k}^-) \cos \theta.
            \label{eq:gv_dimless}
\end{align}
The Faraday term behaves differently because it is a continuum, rather than
a line, effect, and the dimensionless form of eq.(\ref{eq:gammaqu}) is
\begin{equation}
\gamma_{qu,n} = -\frac{4 \Gamma w \nu_c e^3 n_e B \cos \theta}
                      {3\sqrt{\pi}A^0 P^0 \epsilon_0 m_e^2 c^4}
\label{eq:qu_dimless}
\end{equation}

Finally, we note here that we opt not to convert the angular frequencies, $\varpi_n$ and $W$, to dimensionless values, to allow more direct control over selected angular frequency resolution, Doppler, and Zeeman widths in the numerical solver described below.

\section{The numerical solver}
\label{s:code}

We now discuss the numerical implementation of the one-dimensional formalism derived in the previous section. Functionally, the formalism comprised of eq.(\ref{eq:sig_dimless})-(\ref{eq:pi_dimless}) and (\ref{eq:gi_dimless})-(\ref{eq:qu_dimless}) forms a closed system of equations that is solved iteratively to calculate the Stokes parameters and corresponding population inversion for a given set of conditions. 

In practice, according to eq.(\ref{eq:fullformsol}), the solution at any position along a line of sight through the cloud depends on the solution at every other point through which the ray has already traveled. In addition, the unitless inversions (eq.(\ref{eq:sig_dimless}) and (\ref{eq:pi_dimless})) at any given frequency $\varpi_n$ also depend on the solutions at frequencies $\varpi_{n \pm k}$. Therefore, the system of equations must be solved simultaneously at each frequency and optical depth through the masing material. However, they are fully separable along other parameters. 

The one-dimensional radiative transfer solutions and their associated observables were computed with PRISM\footnote{\url{github.com/tltobin/prism}} (Polarized Radiation Intensity from Saturated Masers; \citealt{prism-release}). PRISM, in its current form, was developed from an original code described in \citet{tobin19}. PRISM solves the dimensionless system of equations presented in the previous section simultaneously across a 2-dimensional grid of angular frequency, $\varpi_n$, and optical depth, $\tau$, up to a given total optical depth for the cloud, $\tau_f$. The numerical solution for the dimensionless inversions, $\delta^{\pm,0}$ is derived for each set of parameters using SciPy's \texttt{optimize.newton\_krylov} zero-finder. The residual between the starting $\delta^{\pm,0}$ for each iteration and the resulting value defined by eq.(\ref{eq:sig_dimless}) and (\ref{eq:pi_dimless}) is adjustable in PRISM; however, for all numerical solutions presented here, the tolerance for the solution at each grid point was set at $6 \times 10^{-10}$.

The remaining parameters -- $\theta$, $\phi$, $W$, $\eta^{\pm}$, $\alpha^{\pm}$, $k$, the number of expansion terms used for eq.(\ref{eq:fullformsol}), $M$, and the dimensionless background Stokes parameters, $\vect{i}(0)$ -- are treated as constant throughout the cloud for a given solution. Faraday rotation can be included by either explicitly specifying a non-zero $-\gamma_{qu}/\cos\theta$ or by calculating it from eq.(\ref{eq:qu_dimless}). In addition, PRISM can solve systems with either a single ray travelling in one direction or two opposing rays travelling in opposite directions through the one-dimensional cloud. The modifications to the formalism presented in the previous section that are required for the case of counter-propagating rays are detailed in Appendix \ref{a:bidirectional}. In the case of counter-propagating rays, PRISM does not require the background Stokes parameters for the second ray to be identical to those of the first ray.

In practice, solving the system of equations iteratively at all grid points in $(\varpi_n, \tau)$ space with an $M$-fold expansion of integrated gain matrices can be time consuming, particularly as the total $\tau_f$ increases. To alleviate this concern, solutions are calculated for a given system for gradually increasing total optical depth. The calculation of $\delta^{\pm,0}$ at each subsequent total optical depth uses the $\delta^{\pm,0}$ solution for the previous total optical depth as its initial guess. In the numerical solutions presented here, we begin with a total optical depth of $\tau_f = 0.1$, which typically converges to within tolerance in only a few iterations of the \texttt{newton\_krylov} solver. 

While the method above is built directly into the PRISM code, other techniques may be required to speed convergence, particularly in cases where $d\tau_f > 5 $ between successive iterations. In the case of large differences between $\tau_f$ in successive iterations, intermediate steps in $\tau_f$ with less stringent tolerances for convergence (eg. $1.0$ instead of $6 \times 10^{-10}$) may provide a more accurate initial guess for the next desired $\tau_f$ solution, despite the additional derivations.

Another time-saving option is to derive the solution at all desired $\tau_f$ using the method above at a single $\theta_j$, and then use the $\delta^{\pm,0} (\theta_j)$ solution as the initial guess for deriving the $\delta^{\pm,0} (\theta_{j+1})$ solution, provided that the step between $\theta_j$ and $\theta_{j+1}$ is small enough. With the $\theta$ sampling used in following sections ($d\theta = 2.5^{\circ}$ for $\theta < 30^{\circ}$, $d\theta \sim 5 ^{\circ}$ for $\theta \geq 30^{\circ}$), this method typically provides an improvement of several magnitudes in the initial residual for each new solution, and does not require the addition of intermediate $\tau_f$. However, this method does require solutions at a single $\theta$ to already exist for any desired $\tau_f$. Therefore, the method of choice when performing the work presented here for a set of solutions as a function of $\theta$ is to first compute solutions as a function of increasing $\tau_f$ for $\theta = 0^{\circ}$ and $90^{\circ}$ individually, using intermediate $\tau_f$ as needed. These results are then used as anchors to calculate new solutions for desired $\tau_f$ only as $\theta$ increases from $0^{\circ}$ and decreases from $90^{\circ}$. Once a population solution has been obtained in the
$(\varpi_n , \tau)$ space, computationally cheap formal solutions recover the Stokes parameters from 
eq.(\ref{eq:rteq}) and appropriate radiation boundary
conditions.

When calculating the dimensionless Stokes parameters, \vect{i}, following eq.(\ref{eq:fullformsol}), the number of expansion terms required to converge on a solution to some desired precision increases substantially with $\tau_f$ and $-\gamma_{qu}$. However, the convergence of a solution also depends on other parameters, such as $\theta$, to a lesser extent. The convergence for each solution is evaluated after calculation, to ensure that the desired precision was reached. Convergence is characterized for each full solution via a two-fold analysis of the absolute value of the expansion terms in eq.(\ref{eq:fullformsol}). For each solution grid calculated as a function of angular frequency from line center, $\varpi$, and optical depth through the cloud, $\tau$, we calculate the maximum fractional contribution to each Stokes parameter from the $M$th expansion term across all $(\varpi,\tau)$ bins, or max$( | \Delta \pmb{i} | / \pmb{i} )$. We then verify the results for a converging trend as the number of expansion terms, $M$, increases. The convergence of a suite of solutions across $\theta$ and total optical depth, $\tau_f$, are then characterized according to the maximum fractional contribution to a Stokes parameter for any $\theta$ at the final utilized expansion term for the highest total optical depth. 

\section{Results}
\subsection{Large and Small Zeeman Splitting Solutions}

\begin{figure}[h!]
\begin{center}
\includegraphics[width=0.95\linewidth]{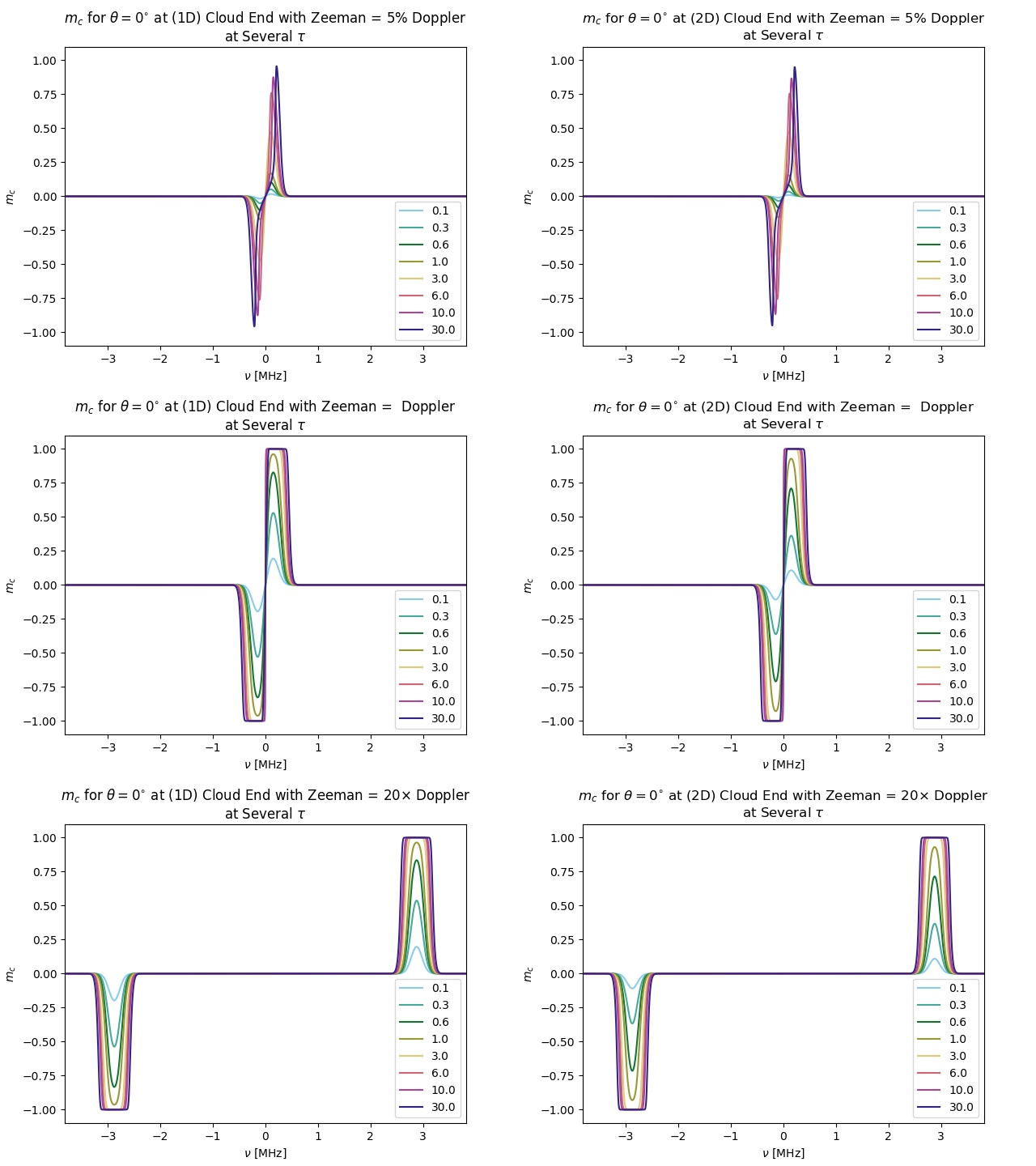}
\caption{Calculated Stokes $v/i$ at the end of a one-directional (left column) and two-directional (right column) cloud with $\theta = 0^{\circ}$ at several values of Zeeman splitting relative to Doppler line width. Each figure shows the resulting profile of $v/i$ with frequency in MHz for a range of total optical depths.}\label{fig:splitting_theta0}
\end{center}
\end{figure}

\begin{figure}[h!]
\begin{center}
\includegraphics[width=0.95\linewidth]{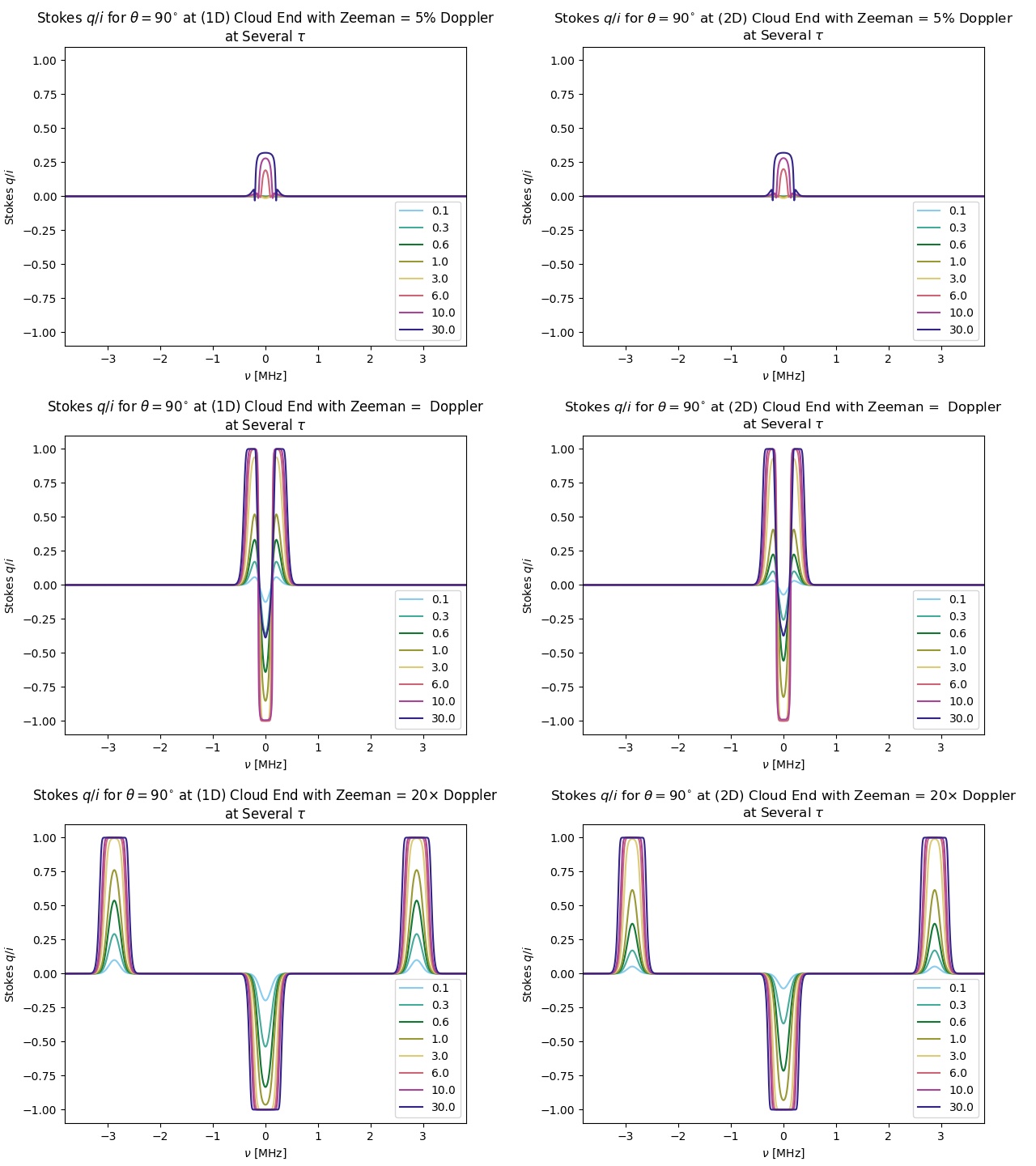}
\caption{The same as Figure \ref{fig:splitting_theta0}, but showing Stokes $q/i$ for $\theta = 90^{\circ}$.}\label{fig:splitting_theta90}
\end{center}
\end{figure}

Although most of the results discussed in this work will focus on small Zeeman splitting at the level of SiO, the lack of assumptions about the strength of Zeeman splitting relative to the Doppler width in deriving the formalism provides a general formulation applicable to different molecular species. To demonstrate this, we derived solutions for the cases of a magnetic field parallel ($\theta=0^{\circ}$) and perpendicular ($\theta=90^{\circ}$) to the line of sight for uni-directional and bi-directional clouds. Using a Doppler width of $W = 9.038 \times 10^5$ s$^{-1}$ (SiO $\nu=1$, $J=1-0$ ; \citealt{2005ApJ...625..978B,2004JPCRD..33..177L}), we vary the Zeeman splitting via $k$, calculating a solution for a Zeeman shift of $\Delta \omega = \lbrace 0.05, 0.25, 0.5, 1, 2, 4, 20 \rbrace \times W$ for total optical depths, $\tau_f$, from $0.1 - 30$. All used $\phi=0$ with $M=50$ expansion terms, for simplicity, and had a 5001 resolution elements along $\varpi$ covering the range $\pm 62.5 W$, with 101 resolution elements along $\tau$. 

The resulting fractional circular polarization ($m_c$) profiles with $\theta=0^{\circ}$ and linear polarization ($m_l$) profiles with $\theta = 90^{\circ}$ at the end of the cloud are shown in Figures \ref{fig:splitting_theta0} and \ref{fig:splitting_theta90}, respectively, for several representative $\Delta \omega / W$. Most notably, for large Zeeman splitting (eg. $\Delta \omega / W = 20$), the polarization profile for $\theta = 0^{\circ}$ and $90^{\circ}$ approaches the solution for the normal longitudinal and transverse Zeeman effect, respectively. In these cases, the center of each component reaches 100\% polarization at a total optical depth, $\tau_f$, between 3 and 6.

\begin{deluxetable}{cll}
    \caption{Parameters Used for SiO Solutions\label{tbl:parameter_summary}}
    \tablehead{
        \colhead{Parameter} & \colhead{Description} & \colhead{Value}
    }
    \startdata
        $\nu_c$ & Line Center Frequency & 43.122 GHz \\
        $W$ & Doppler Width [Angular Frequency] & $9.038 \times 10^5$ s$^{-1}$ \\
        $w$ & Doppler Width [velocity]\tablenotemark{a} & 997.8 m s$^{-1}$ \\
        $\Delta \omega$ & Single Substate Zeeman Splitting [Angular Frequency] & 740.48 $\times $ ( B / 1 G ) s$^{-1}$ \\
        $\phi$ & Sky-plane angle & $0^{\circ}$ \\
        $\eta^{\pm}$ & Squared Substate Dipole Moment Ratio & 1\tablenotemark{g} \\
        $\alpha^{\pm}$ & Substate Pumping Rate Ratio & 1 \\
        $\vect{i}_0$ & Initial Unitless Stokes ($i_0$,$q_0$,$u_0$,$v_0$)\tablenotemark{b} & ($10^{-8.2}$, 0, 0, 0) \\
        $\Gamma$ & Loss Rate\tablenotemark{a} & 5 s$^{-1}$ \\
        $A^0$ & Einstein A Coefficient\tablenotemark{a} & $3 \times 10^{-6}$ s$^{-1}$ \\
        $P^0$ & Pump Rate per Volume into 0 Substate\tablenotemark{a} & $1.5 \times 10^6$ cm$^{-3}$ s$^{-1}$ \\
    \hline
        $\varpi$ & Sampled Angular Frequency & $\left\lbrace -2500, -2499, ..., 2500 \right\rbrace \times$ 740.48 s$^{-1}$\\
        $\Delta \tau$ & L.o.S. Optical Depth Resolution & $\left\lbrace 0, 0.01, ..., 1 \right\rbrace \times \tau_f$ \\
        $\theta$ & Angle between $\vect{B}$ and L.o.S. & $\left\lbrace 0, 2.5, ..., 30, 35, ..., 50, 54, 56, 60, 65, ..., 90 \right\rbrace ^{\circ}$ \\
        $B$ & Magnetic Field Strength & $\left\lbrace 1, 2, 5, 10 \right\rbrace$ G\tablenotemark{c} \\
        $k$ & Zeeman Splitting in Ang. Freq. Bins & $\left\lbrace 1, 2, 5, 10 \right\rbrace$\tablenotemark{c} \\
        $n_e$ & Electron Number Density\tablenotemark{d} & $\left\lbrace 0, 7.85e3, 1.57e5 \right\rbrace$ cm$^{-3,}$\tablenotemark{e} \\
        $\tau_f$ & Total Optical Depth & $\left\lbrace 0.1, 0.3, 0.6, 1, 1.1, ..., 3, 4.5, 6, 8, 10, 13, 16, 20, 25, 30, 45, 60, 80, 100 \right\rbrace$\tablenotemark{f} \\
        $M$ & Number of Expansion Terms & 50 for $n_e = 0$ cm$^{-3}$ or ( B = 1~G, $n_e = 7.85e3 \textrm{ cm}^{-3}$) \\
        & & 70 for ( B = 5~G, $n_e = 7.85e3 \textrm{ cm}^{-3}$) \\
        & & 80 for ( B = 10~G, $n_e = 7.85e3 \textrm{ cm}^{-3}$) \\
        & & 100 for ( B = 1~G, $n_e = 1.57e5 \textrm{ cm}^{-3}$) \\
        & & 110 for ( B = 5~G, $n_e = 1.57e5 \textrm{ cm}^{-3}$) \\
        & & 120 for ( B = 10~G, $n_e = 1.57e5 \textrm{ cm}^{-3}$)
    \enddata
    \tablenotetext{a}{Only used for cases with nonzero Faraday rotation $(\gamma_{qu} \neq 0)$. }
    \tablenotetext{b}{For bi-directional integration, the same initial Stokes are used for each ray.}
    \tablenotetext{c}{B $=$ 2 G $(k=2)$ only used for uni-directional integration with no Faraday Rotation ($n_e = 0$ cm$^{-3}$).}
    \tablenotetext{d}{Used to turn on/off Faraday Rotation. An $n_e = 0$ cm$^{-3}$ yields a $\gamma_{qu} = 0$. }
    \tablenotetext{e}{Solutions only computed with $n_e = 1.57e5$ cm$^{-3}$ for uni-directional clouds.}
    \tablenotetext{f}{Solutions computed with $n_e = 1.57e5$ cm$^{-3}$ are only integrated up to maximum $\tau_f = \left\lbrace 30,16,10 \right\rbrace$ for B $= \left\lbrace 1, 5, 10 \right\rbrace$ G, respectively.}
    \tablenotetext{g}{This must be 1.0 for the present system; we leave it as a parameter to allow for 
    future work with more complicated Zeeman patterns.}
\end{deluxetable}

\subsection{SiO Polarization with No Faraday Rotation\label{ss:siopolnof}}

\begin{figure}[h!]
\begin{center}
\includegraphics[width=0.9\linewidth]{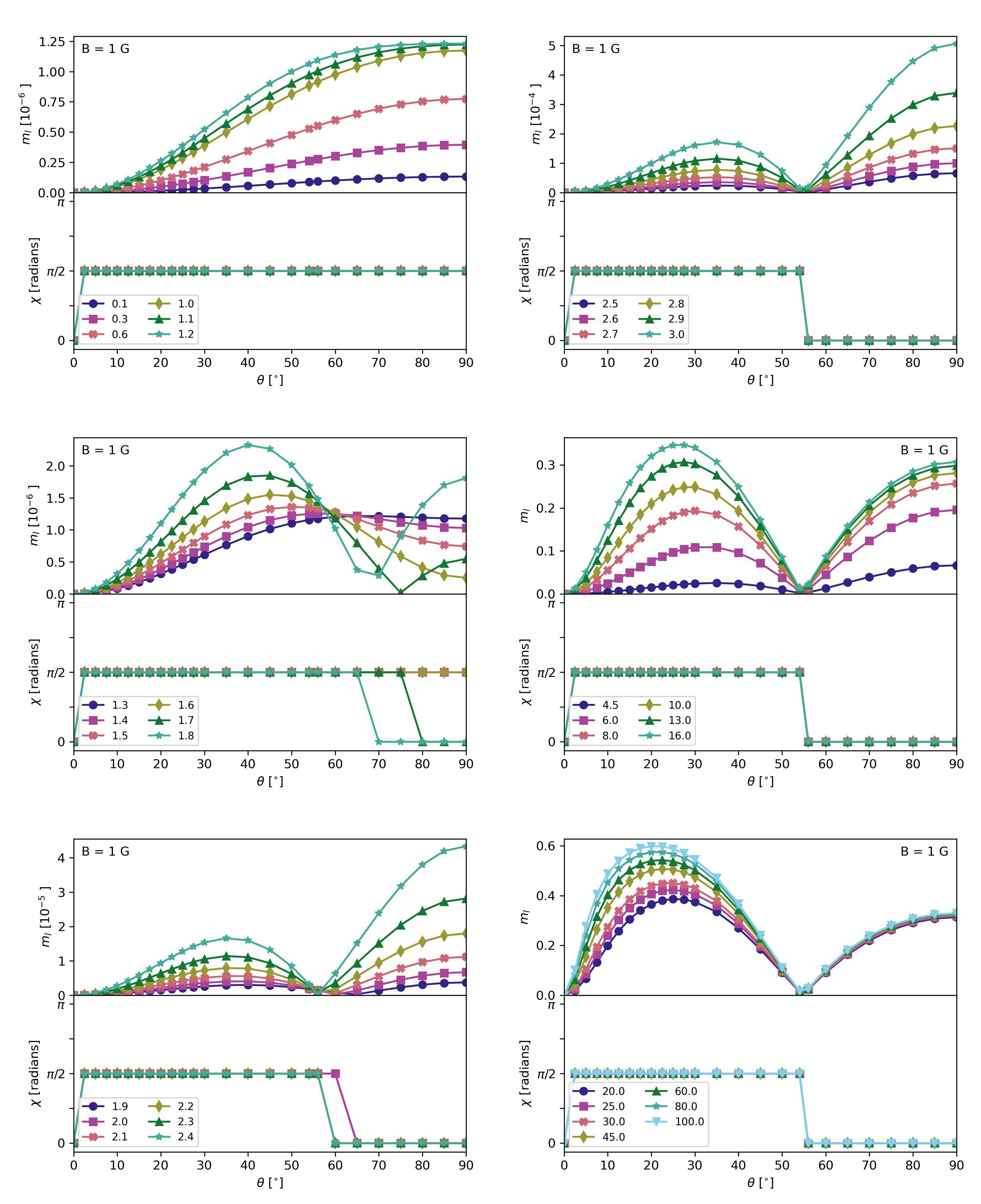}
\caption{Fractional linear polarization, $m_l$, and EVPA, $\chi$, at line center as a function of $\theta$ for a one-directional cloud with a 1~G magnetic field and no Faraday rotation. Each curve corresponds to a different total optical depth, $\tau_f$, denoted in the legend. Curves are split into multiple plots for clarity, with $\tau_f$ increasing first downward within a column and then across the row of figures. }\label{fig:mlevpa_v_theta_nofr_k01}
\end{center}
\end{figure}

We consider next the case of the SiO $\nu = 1$, $J = 1-0$ transition as a function of $\theta$ for a range of total optical depths, $\tau_f$, and magnetic field strengths. For the moment, we ignore Faraday Rotation by setting $\gamma_{qu} = 0$. We again use a Doppler width of $W = 9.038 \times 10^5 s^{-1}$ with a line center frequency of 43.122 GHz \citep{2005ApJ...625..978B,2004JPCRD..33..177L}. The angular frequency shift with respect to line center due to Zeeman splitting is given by $\Delta \omega = \frac{1}{2} g \Omega = ( 740.48 \textrm{ s}^{-1}\textrm{ G}^{-1} ) \times B$ \citep{2013A&A...551A..15P}. The latter is implemented by setting the angular frequency array as $\varpi = \left\lbrace -2500, -2499, ... 2500 \right\rbrace \times 740.48 \textrm{ s}^{-1}$, and selecting $k = B$ [G]. As before, we use a resolution of 101 elements along $\tau$ for each calculation and 50 expansion terms. We also set $\phi = 0^{\circ}$ for simplicity.

We compute a suite of uni-directional solutions in the cases of $k =$ 1, 2, 5, and 10, corresponding to magnetic field strengths of $B = $ 1~G, 2~G, 5~G, and 10~G, respectively. Within each solution, $\theta$ varies from $0^{\circ} - 90^{\circ}$ and $\tau$ ranges from $0.1 - 100$. Within this grid, $\theta$ increases in steps of $2.5^\circ$ through $\theta = 30^\circ$, followed by steps of $5^{\circ}$ thereafter; the exception is the $55^{\circ}$ solution, which is replaced by solutions at $54^{\circ}$ and $56^{\circ}$ for increased resolution around the Van Vleck angle (GKK). Solutions were computed for optical depths of $\tau_f = \lbrace$ 0.1, 0.3, 0.6, (1.0 - 3.0 in steps of 0.1), 4.5, 6, 8, 10, 13, 16, 20, 25, 30, 45, 60, 80, 100 $\rbrace$. We compute a similar set of bi\-directional solutions for 1~G, 5~G, and 10~G magnetic fields. We used $M=50$ expansion terms in all cases, achieving a precision of max$( | \Delta \pmb{i} | / \pmb{i} ) \lesssim 2 \times 10^{-6}$ at all sampled optical depths. A summary of the parameters used to derive the solutions described in this and following sections is shown in Table \ref{tbl:parameter_summary}.

The behavior of the scaled intensity, $i = I/I_{sat}$, is similar across all cases. For a given optical depth and number of rays, $i$ changes little from $\theta = 90^{\circ}$ - $40^{\circ}$, decreasing slightly ($\sim 10 \%$) as $\theta$ approaches $0^{\circ}$. While there is no significant change in $i$ with $B$, increasing the number of rays from one to two decreases the total $i$ by about a factor of 2. In the cases discussed here, the maximum $i$ reached for $\tau = 100$ is $\sim 71$ for a uni-directional maser and $\sim 35$ for a bi-directional maser.

The resulting fractional linear polarization ($m_l$) and its position angle (EVPA) at line center for a one-directional maser in a 1~G magnetic field are shown in Figure \ref{fig:mlevpa_v_theta_nofr_k01} as a function of $\theta$ for all sampled $\tau_f$. At this magnetic field strength, $m_l$ is low ($ < 1.5 \times 10^{-6}$) for $\tau_f$ up to $\sim 1.2$, increasing for larger $\theta$ and showing no flip in EVPA. We show in Appendix~\ref{appendix3} that no flip is expected for negligible saturation. As $\tau_f$ increases further ($\tau \sim 1.3 - 1.6$), $m_l$ at higher $\theta$ begins to decrease even as it continues to increase at $\theta \sim 40^{\circ}$, forming a singularly-peaked function. 

The EVPA flip appears first at $\tau_f \sim 1.7$, but at $\theta \sim 75^{\circ}$, while $m_l$ at $\theta$ larger than the flip location increases once again to form half of a second peak. As $\tau_f$ continues to increase up to $\sim 2.4$, the $\theta$ at which the EVPA flip occurs approaches its final location between $\theta = 54^{\circ}$ and $56^{\circ}$; meanwhile, $m_l$ increases in both peaks, though $m_l$ at $\theta = 90^{\circ}$ quickly surpasses the peak at more moderate $\theta$. 

By $\tau_f \sim 2.5$, the EVPA flip has reached the Van Vleck angle to our $\theta$ resolution, where it stays for all remaining $\tau_f$. The $m_l$ in both peaks continues to grow, with the peak at moderate $\theta$ finally reaching and surpassing that at $\theta = 90^{\circ}$ around $\tau_f \sim 13$, when $m_l \sim 0.3$ at both peaks. From there, $m_l$ continues to increase in the lower $\theta$ peak, with only slight increases in the $m_l$ of the $\theta = 90^{\circ}$ peak.

\begin{figure}[h!]
\begin{center}
    \includegraphics[width=0.92\linewidth]{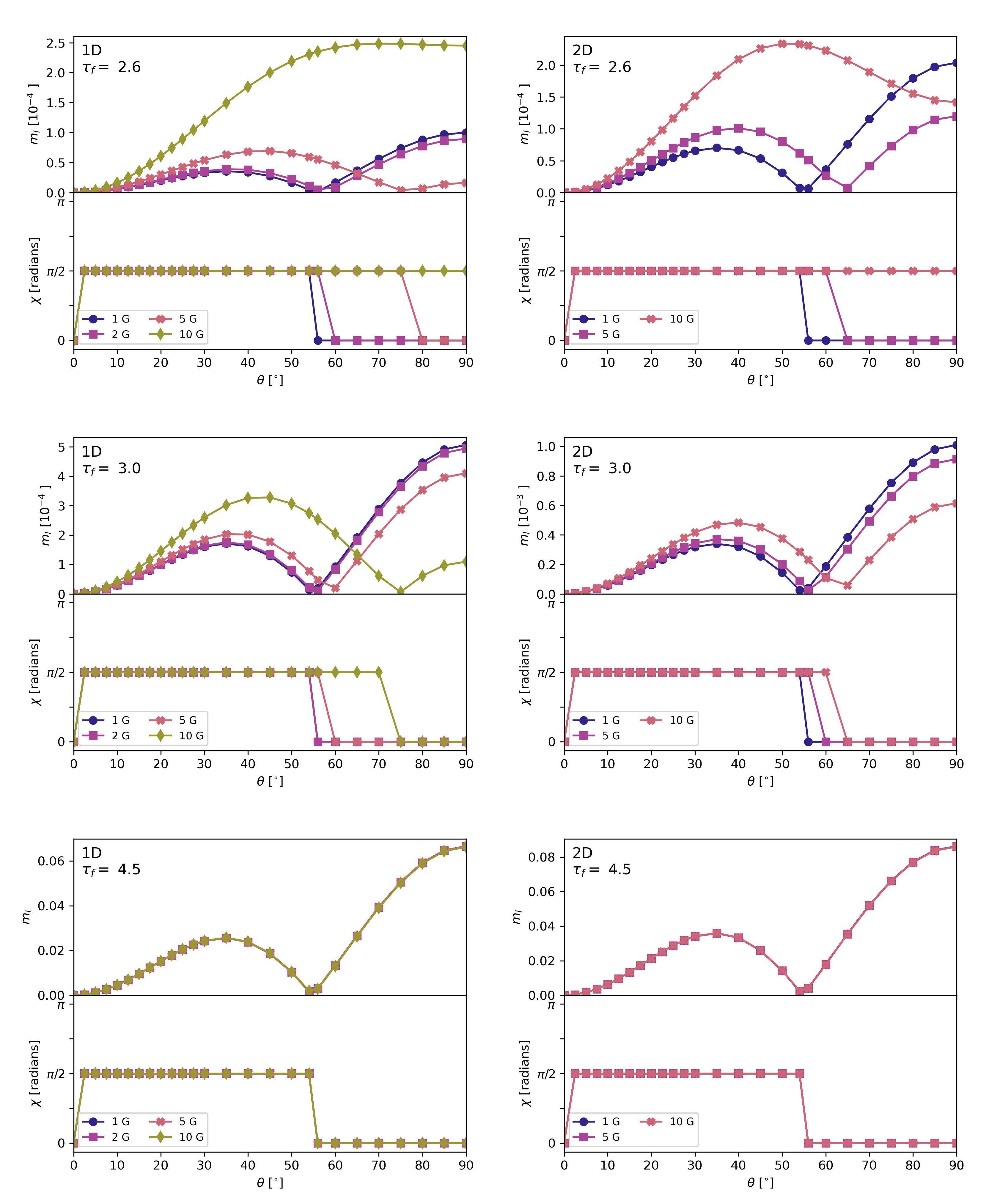}
    \caption{Fractional linear polarization, $m_l$, and EVPA, $\chi$, at line center as a function of $\theta$ for uni-directional \textit{(left column)} and bi-directional \textit{(right column)} clouds with no Faraday rotation for total optical depths of $\tau_f = 2.6$ \textit{(top)}, 3.0 \textit{(middle)}, and 4.5 \textit{(bottom)}. The magnetic field strength of each curve is denoted in the legend, showing the variation in the $\tau_f$ at which the EVPA flip sets in as the magnetic field strength increases.}\label{fig:mlevpa_v_theta_nofr}
\end{center}
\end{figure}

As shown in Figure \ref{fig:mlevpa_v_theta_nofr}, increasing the magnetic field strength increases both the scale of the fractional linear polarization present and the total optical depth, $\tau_f$, required for the appearance of the EVPA flip and its approach to the Van Vleck angle. A comparison to the bi-directional cases shows that increasing from one to two rays causes the overall scale of $m_l$ to increase more quickly with $\tau_f$, in addition to requiring lower $\tau_f$ for the appearance of the EVPA flip and its approach of the Van Vleck angle for a given magnetic field strength. 

However, by $\tau_f = 4.5$, the EVPA flip occurs between $\theta = 54^{\circ}$ and $56^{\circ}$ in all cases and $m_l$ no longer varies significantly with B. For an optical depth of $\tau_f=100$, the largest difference in $m_l(\theta)$ at line center between a 1~G and a 10~G magnetic field is $\sim7 \times 10^{-5}$ for one-directional propagation and $\sim 6 \times 10^{-5}$ for bi-directional propagation. Compared to the peak $m_l(\theta)$ values of 0.60 and 0.74, respectively, the variation from magnetic field strength alone is not visible to the eye when plotting $m_l(\theta)$ at these high optical depths.

\begin{figure}[h!]
\begin{center}
    \includegraphics[width=0.95\linewidth]{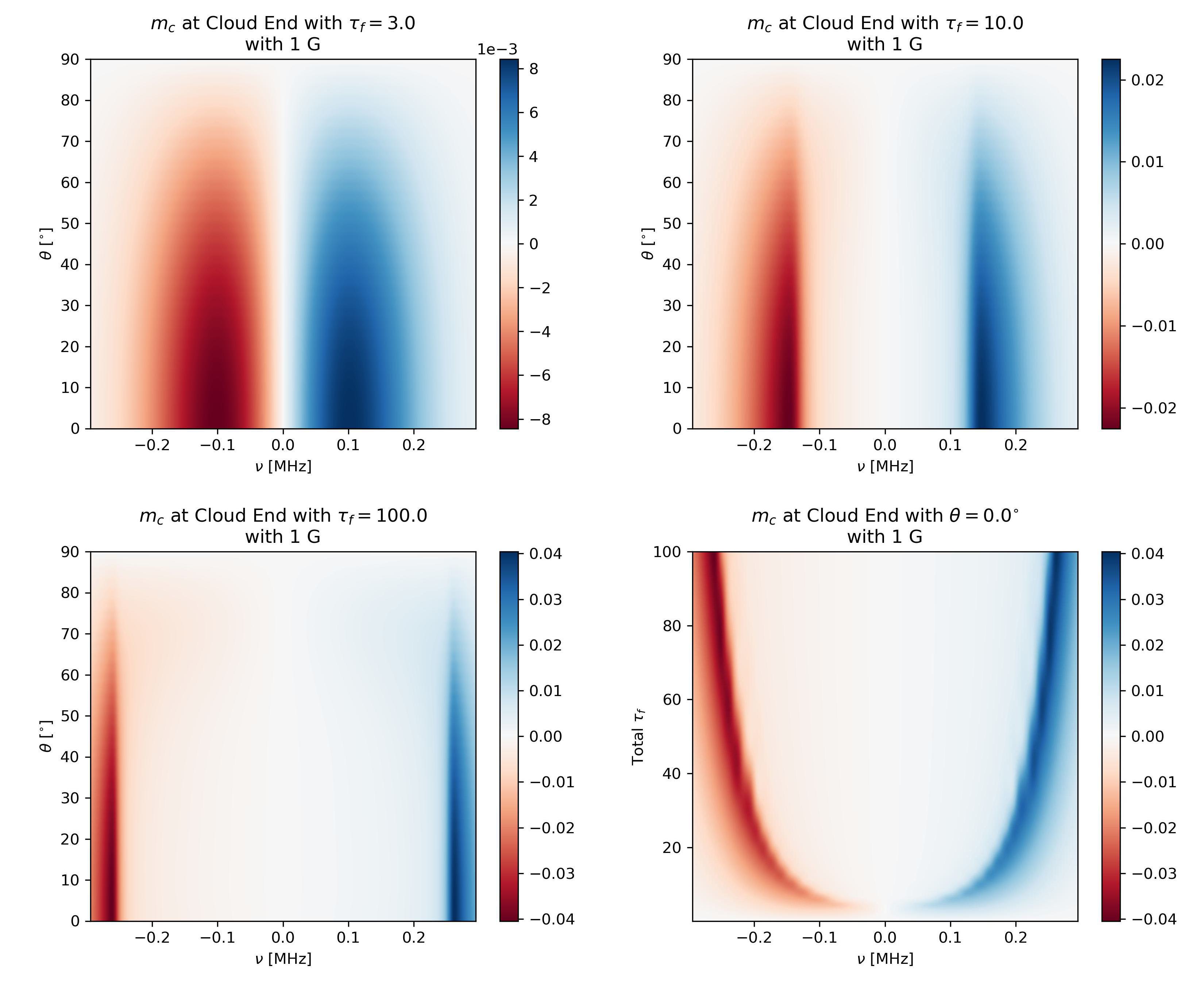}
    \caption{Fractional circular polarization ($m_c$) at the cloud end for a uni\-directional maser with no Faraday Rotation and a 1~G magnetic field. First three figures show $m_c$ as a function of frequency, $\nu$, and angle $\theta$ for clouds with total optical depth of $\tau_f=3$ (\textit{upper left}), 10 (\textit{upper right}), and 100 (\textit{lower left}). \textit{Lower right} figure shows $m_c$ as a function of frequency, $\nu$, and total optical depth, $\tau_f$, for a magnetic field parallel to the line of sight. } \label{fig:mc_grid_1D}
\end{center}
\end{figure}

\begin{figure}[h!]
\begin{center}
    \includegraphics[width=0.95\linewidth]{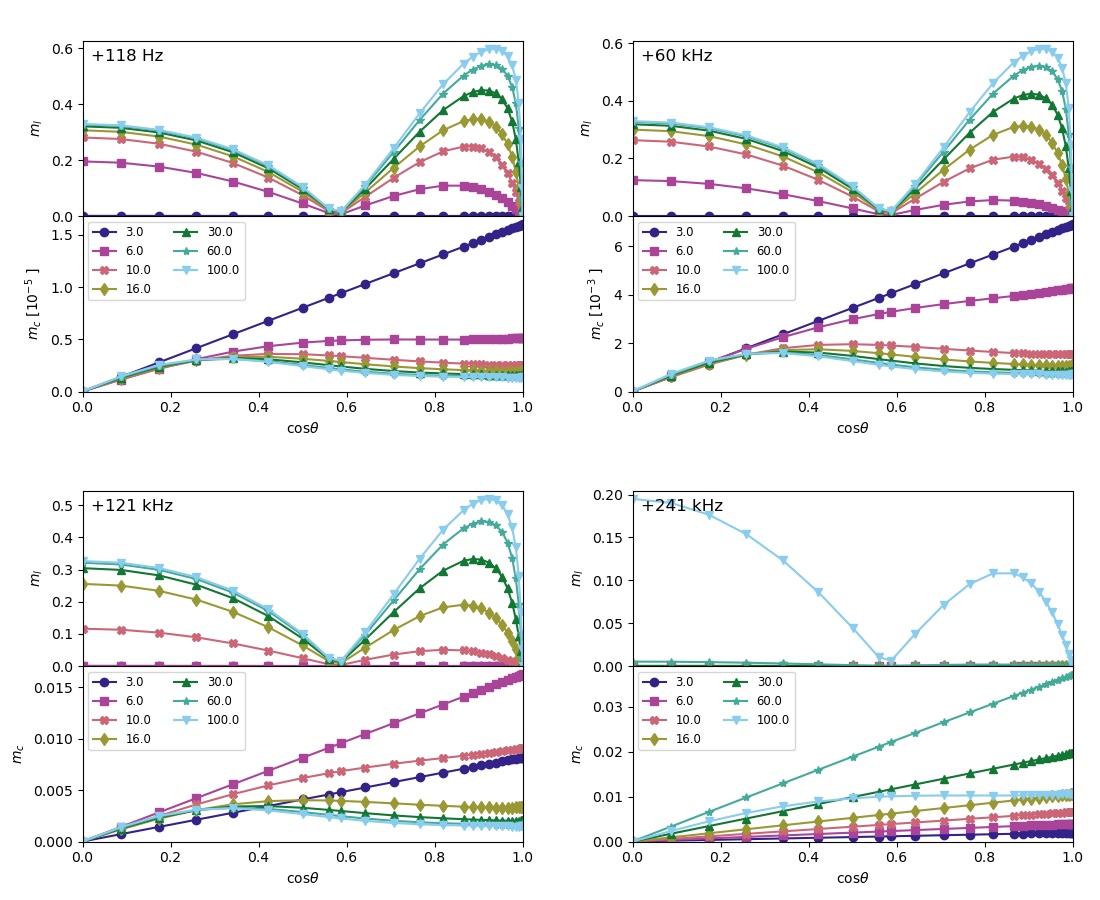}
    \caption{Fractional linear ($m_l$) and circular ($m_c$) polarization at line end as a function of $\cos \theta$ for a uni\-directional maser with no Faraday Rotation and a 1~G magnetic field. As there is no circular polarization at line center, plots are shown at frequency ($\varpi$ bin) offsets of $+118$~Hz (1), $+60$~kHz (512), $+121$~kHz (1024), and $+241$~kHz (2048) from line center. Curves within each plot show the profile at a given total optical depth, $\tau_f$, as denoted in the legend.} \label{fig:mlmc_v_costheta}
\end{center}
\end{figure}

The circular polarization profile is anti\-symmetric about line center, resulting in $m_c = 0$ at line center itself (see Figure \ref{fig:mc_grid_1D}). The maximum $m_c(\nu)$ for a system at cloud end increases for larger values of $\theta$ and $\tau_f$. For $\tau_f = 0.1 - 3.0$, the frequency at which the $m_c$ extrema occur ($\nu \sim \pm 101$ kHz or $\varpi \sim \pm 0.7W$) is constant with $\tau_f$ and $\theta$ within the precision of our frequency resolution. For $\tau_f > 3$, the offset frequencies of the $m_c$ extrema increase approximately logarithmically with $\tau_f$, reaching $\varpi \sim \pm 1.8 W$ ($\nu \sim 260$ kHz) for $\tau_f = 100$. These higher optical depths also show a smaller dependence on $\theta$, with the $m_c$ extrema for $\theta = 0^{\circ}$ occurring $\sim 8150$ s$^{-1}$ ($\sim 0.009 W$ ) further from line center at $\tau_f = 100$ than the extrema with $\theta = 90^{\circ}$. However, this amounts to a $< 0.5\%$ change in frequency across all $\theta$ at a given optical depth, indicating that this effect is several orders of magnitude weaker than the change in $m_c$ peak frequency with $\tau_f$.

Figure \ref{fig:mlmc_v_costheta} shows the $m_c$ and corresponding $m_l$ as a function of $\cos \theta$ at +118 Hz, +60 kHz, +121 kHz, and +241 kHz from line center for $\tau_f$ between 3 and 100. In all cases, $m_c = 0$ at $\cos \theta = 0$; i.e. no circular polarization arises when the magnetic field is oriented perpendicular to the line of sight. For $\tau_f \leq 3$, $m_c \propto \cos \theta$ for all $\varpi \neq 0$, with $\partial m_c / \partial \cos \theta$ increasing with $\tau_f$ for a given $\varpi$. For $\tau_f > 3$, $m_c$ is only linear with $\cos \theta$ at or outside of the $m_c$ peak frequencies. For frequencies interior to the $m_c$ peaks, $m_c$ deviates from linearity with $\cos \theta$, showing a greater decrease in $m_c$ at larger $\cos \theta$. 

\begin{figure}[h!]
\begin{center}
    \includegraphics[width=0.5\linewidth]{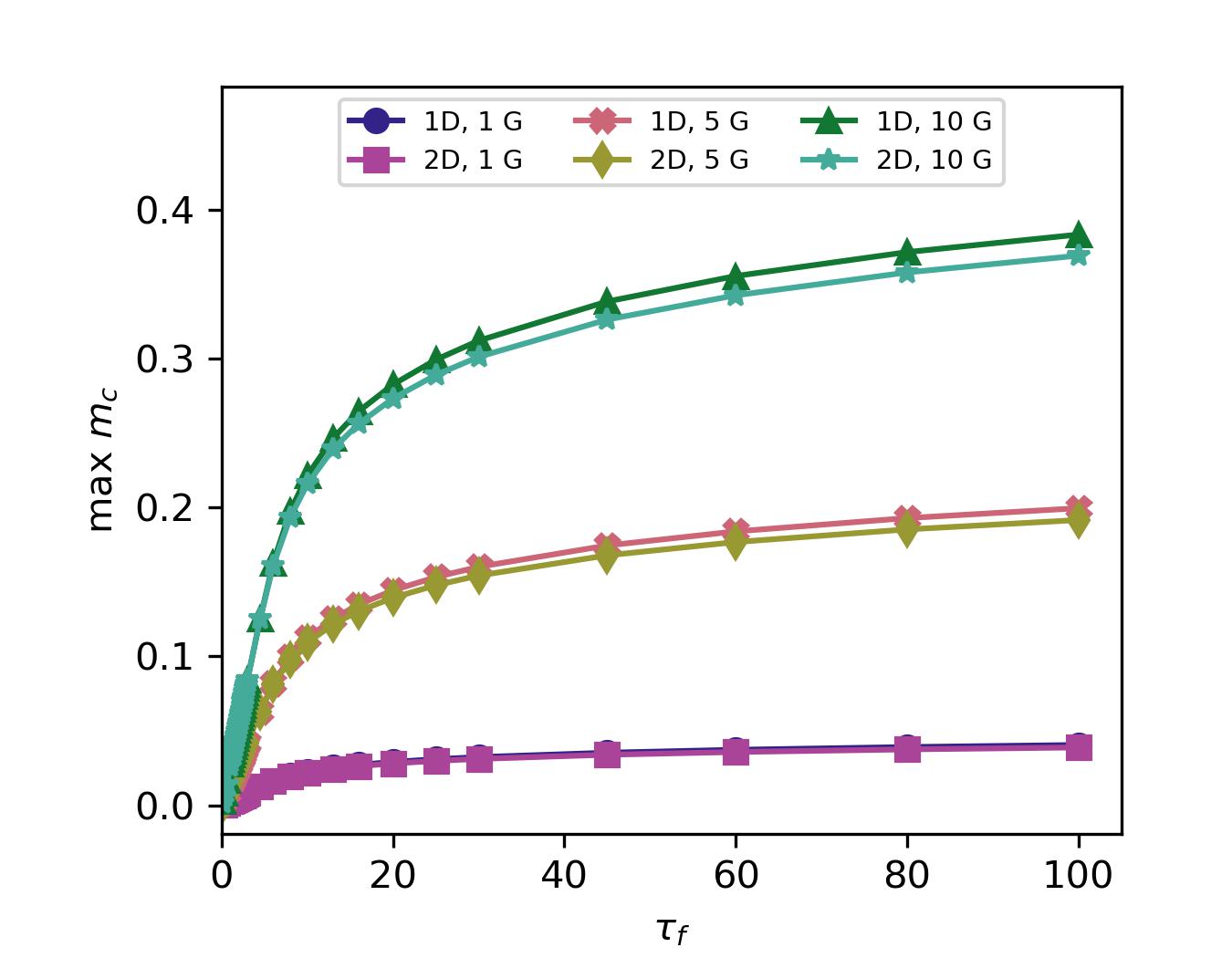}
    \caption{The maximum fractional circular polarization, $m_c$, present in a frequency bin (at the cloud end) as a function of total optical depth, $\tau_f$, with no Faraday rotation. Curves are shown for uni- and bi-directional propagation with 1~G, 5~G, and 10~G magnetic fields.} \label{fig:maxmc_vs_tau}
\end{center}
\end{figure}

A uni-directional cloud with a 1~G magnetic field reaches a maximum fractional circular polarization of $m_c \sim 0.04$ by $\tau_f = 100$ ($\theta = 0^{\circ}$).  It requires $\tau_f > 3$ to achieve a peak $m_c \geq 0.01$ and a $\tau_f > 8$ for a peak $m_c \geq 0.02$ with our frequency resolution. Figure \ref{fig:maxmc_vs_tau} shows how the increase in maximum $m_c$ with $\tau_f$ varies with magnetic field strength and number of propagating rays. The increase in maximum $m_c$ with magnetic field strength is nearly proportional to the magnetic field strength, with a $< 10\%$ deviation from proportionality for $\tau_f$ up to 100. Calculating propagation with bi-directional rays instead of a single ray causes a peak $m_{c,2D} \leq m_{c,1D}$ for a given magnetic field, with values leveling out at $m_{c,2D} \sim 0.96 m_{c,1D}$ for $\tau_f \gtrsim 20$. However, notably, as seen in Figure \ref{fig:mlmc_v_costheta}, the larger frequency offsets from line center that provide $m_c$ of one to a few percent have lower linear polarization than at frequencies close to line center.

\subsection{SiO polarization with nonzero Faraday rotation}

We next consider similar cases with non-zero Faraday Rotation as set by eq.(\ref{eq:qu_dimless}). We maintain the line center frequency, $\nu_c$, of 43.122 GHz for the SiO $\nu=1$, $J=1-0$ transition. The Doppler width in velocity space, $w$, is calculated explicitly as $w = Wc / 2 \pi \nu_c$, for consistency with the set Doppler width $W = 9.038 \times 10^5$ s$^{-1}$. The decay or loss rate, $\Gamma$, is estimated at $\Gamma \sim 5$ s$^{-1}$ \citep{1974ApJ...194L..97K,1990ApJ...354..660N,elitzurbook}. The Einstein-A coefficient, $A^0$, for the $^{28}$SiO $\nu=1$, $J=1-0$ line is set at $3.0490 \times 10^{-6}$ s$^{-1}$ \citep{2005A&A...432..369S}, and the pump rate (per volume) into the 0 substate, $P^0$, is derived from $P^0 = n_{SiO}R$, where $n_{SiO}$ is the number density of SiO molecules and $R$ is the overall pump rate. For an $n_{SiO} \sim 10^5$ cm$^{-3}$ \citep{elitzurbook} and $R \sim 15$ s$^{-1}$ \citep{2013MNRAS.431.1077A} in the SiO masing environments around late-type evolved stars, we estimate a pump rate per volume of $P^0 \sim 1.5 \times 10^6$ cm$^{-3}$ s$^{-1}$.

The electron number density, $n_e$, in the near circumstellar environments of late-type evolved stars is not strongly constrained and can vary significantly within the region. While it can be derived simply from the number density of Hydrogen, $n_H$, and the ionization fraction, $f_{ion}$, estimates of the ionization fraction in SiO masing regions around AGB stars range from $f_{ion} \sim 10^{-7.6}$ \citep{1997ApJ...476..327R} to $f_{ion} \sim 10^{-5}$ \citep{2004agbs.book..149G}. The models of \citet{2011MNRAS.418..114I} have a mean $\log ( \rho [\textrm{g cm}^{-3}]) \sim -13$ at $2.5 \times R_{phot}$, resulting in $n_H \sim 4 \times 10^{10}$ cm$^{-3}$, assuming the relative H$_2$/He abundances from \citet{2016A&A...590A.127W}; however, the precise value of $n_H$ can very by 2 orders of magnitude in either direction depending on the modelled system and sampling time. This is consistent with the upper limit of $n_H < 10^{12}$ cm$^{-3}$ from \citet{2016A&A...590A.127W}. Combined with the range of estimated $f_{ion}$ above, this estimate of $n_H$ would indicate typical values of $n_e$ between $1.3 \times 10^3$ cm$^{-3}$ and $5 \times 10^5$ cm$^{-3}$. 

We compute a grid of unidirectional solutions for each combination of  $n_e = \lbrace  7.85e3, 1.57e5  \rbrace$ cm$^{-3}$ and $B = \lbrace$ 1, 5, 10 $\rbrace$ G. This, combined with the remaining estimated values above, gives us sampled $- \gamma_{qu} / \cos \theta = \lbrace$~0.0193, 0.0965, 0.193, 0.386, 1.93, 3.86~$ \rbrace$. We also calculate the bi-directional solutions for the $n_e =$ 7.85e3 cm$^{-3}$ cases. 

Increasing the Faraday Rotation via $\gamma_{qu}$ requires an increase in the number of expansion terms required to achieve the desired precision. The number of expansion terms, ranging from $M=50 - 120$, are shown in Table \ref{tbl:parameter_summary}. Solutions calculated with $n_e =$ 7.85e3 cm$^{-3}$ reach a precision of max$( | \Delta \pmb{i} | / \pmb{i} ) \lesssim 1 \times 10^{-5}$ for uni-directional clouds and $\lesssim 2 \times 10^{-6}$ for bi-directional clouds out to a total optical depth of 100. Solutions using $n_e =$ 7.85e3 cm$^{-3}$ were only computed up to total optical depths of $\tau_f = \left\lbrace 30, 16, 10 \right\rbrace$ for $ B = \left\lbrace 1, 5, 10 \right\rbrace$ G, respectively. Although convergences at these maximum optical depth reached max$( | \Delta \pmb{i} | / \pmb{i} ) \lesssim 7 \times 10^{-6}$, higher total optical depths resulted in either a significantly degraded convergence evaluated from the expansion term contributions or a failure to converge within the Newton Krylov solver. This behavior may be due to the increased Faraday Rotation generating structure in the Stokes parameters across the $(\varpi,\tau)$ grid that is too fine for our grid resolution to sample smoothly.

\begin{figure}[h!]
\begin{center}
    \includegraphics[width=0.92\linewidth]{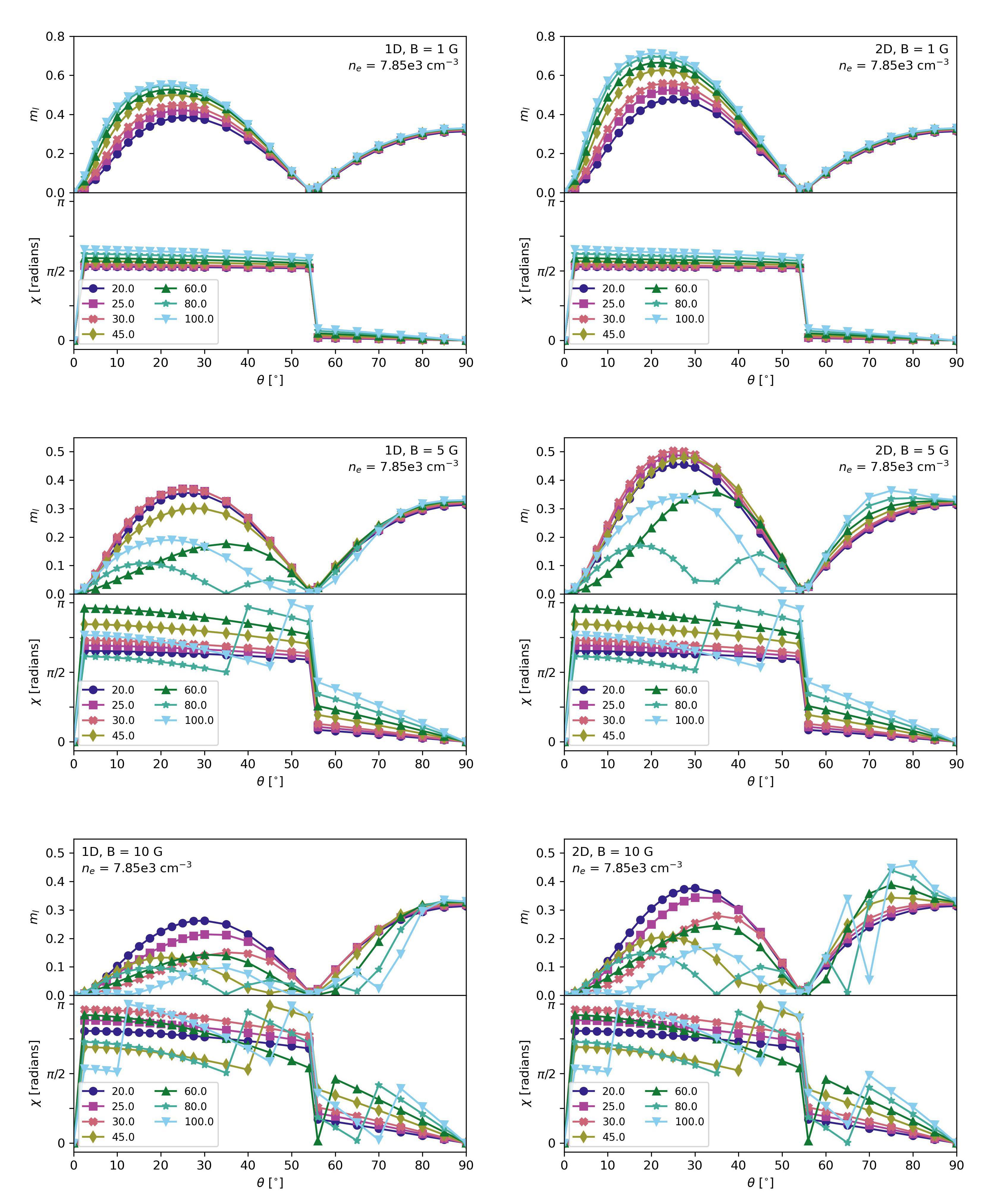}
    \caption{Fractional linear polarization, $m_l$, and EVPA, $\chi$, at line center as a function of $\theta$ with $n_e = $ 7.85e3 cm$^{-3}$ for total optical depths from $\tau_f = 20 - 100$. Figures are shown for magnetic field strengths of 1~G (\textit{top row}), 5~G (\textit{middle row}), and 10~G (\textit{bottom row}) for uni-directional (\textit{left column}) and bi-directional (\textit{right column}) integration.} \label{fig:mlevpa_v_theta_fcA_hightau}
\end{center}
\end{figure}

While behavior of the total intensity, $i$, and fractional circular polarization, $m_c$, do not change significantly compared to the solutions with no Faraday Rotation, the effects of Faraday Rotation on linear polarization can be seen for $n_e = 7.85e3$ cm$^{-3}$ at the highest sampled optical depths (Figure \ref{fig:mlevpa_v_theta_fcA_hightau}). For the weaker Faraday Rotation seen with 1~G magnetic fields, this manifests itself as a slight decrease in the amplitude of the fractional linear polarization and a slight rotation in EVPA as a function of $\theta$, though the EVPA flip and the general $m_l(\theta)$ profile is preserved. However, further increasing the magnitude of the Faraday Rotation can instigate additional EVPA flips. For a 5~G magnetic field, a secondary flip only occurs for optical depths $> 60$; however, with 10~G magnetic fields, we see as many as three additional EVPA flips occurring by an optical depth of 100, with the first appearing at an optical depth of 45. The increasing number of minima in $m_l(\theta)$ that accompanies the appearance of new EVPA flips, also has the effect of further decreasing the maximum $m_l(\theta)$ that can be present at line center.

Even for a given electron density and magnetic field strength, the Faraday Rotation term $\gamma_{qu} \propto \cos \theta$. As expected, no change in linear polarization is present for magnetic fields perpendicular to the line of sight ($\theta = 90^{\circ}$), with the effect of Faraday Rotation generally more apparent as $\theta$ decreases. We also reproduce the expected non-reciprocity of Faraday Rotation; for bi-directional clouds, the Stokes solution is symmetric about cloud center.

\begin{figure}[h!]
\begin{center}
    \includegraphics[width=0.92\linewidth]{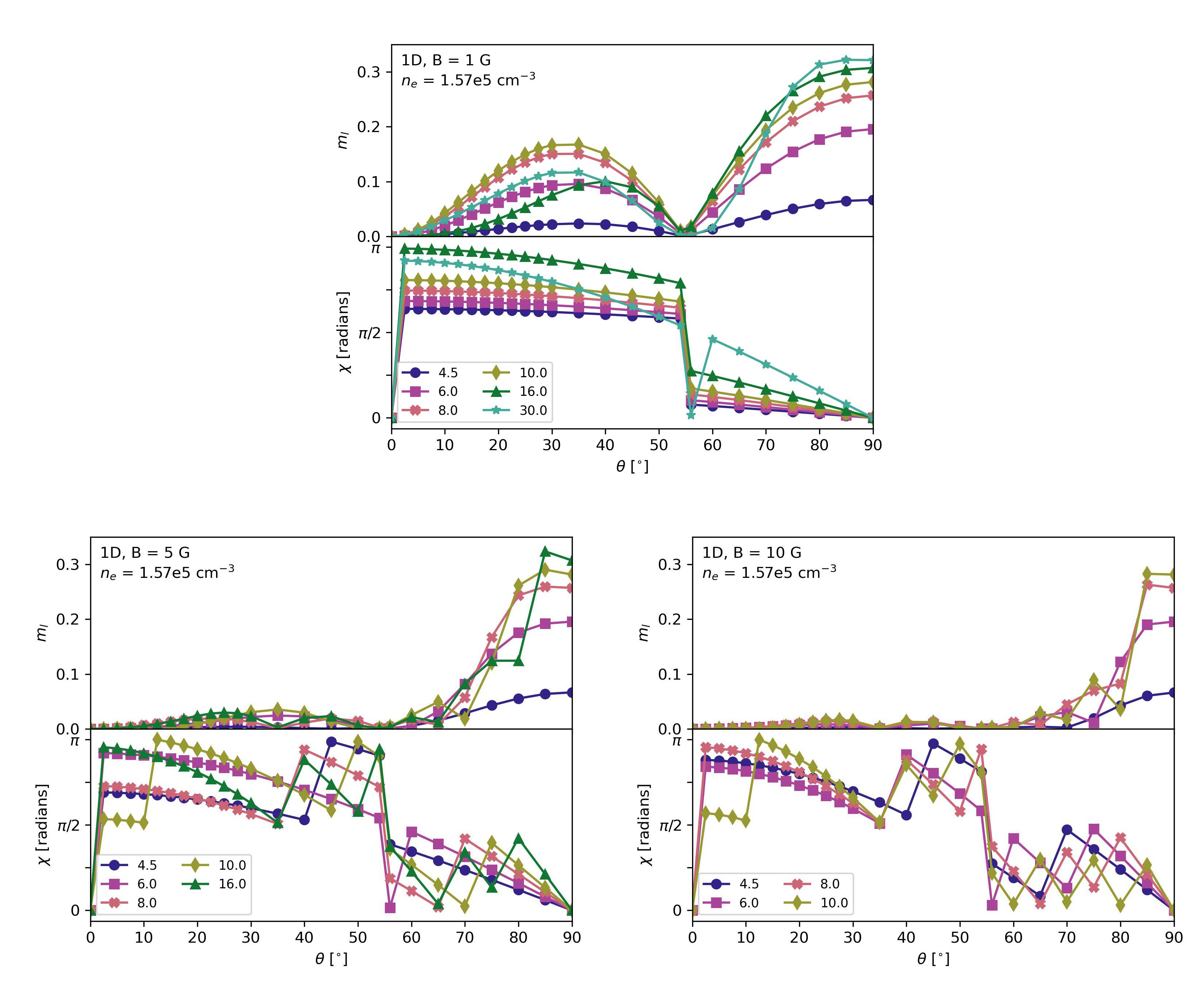}
    \caption{Fractional linear polarization and EVPA at line center as a function of $\theta$ for uni-directional integration with $n_e = $1.57e5~cm$^{-3}$. Figures are shown for 1~G (\textit{top}), 5~G (\textit{lower left}), and 10~G (\textit{lower right}) magnetic fields, with the total optical depth indicated in the legend.} \label{fig:mlevpa_v_theta_fcB_hightau}
\end{center}
\end{figure}

The linear polarization for cases with a higher electron density of $n_e = 1.57e5$ cm$^{-3}$ is shown in Figure \ref{fig:mlevpa_v_theta_fcB_hightau} up to the maximum total optical depth achieved for each sampled magnetic field strength. With a 1~G magnetic field ($-\gamma_{qu} / \cos \theta = 0.386$), the appearance of secondary EVPA flips begins for $\tau_f > 16$. The fractional linear polarization at lower $\theta$, while still showing a smooth variation with $\theta$, is suppressed enough that it never exceeds the fractional linear polarization at $\theta = 90^{\circ}$ once the Van Vleck angle has reached its characteristic value. 

Further increasing the Faraday Rotation to $-\gamma_{qu} / \cos \theta = 1.93$ ($B=$ 5~G) or 3.86 ($B=$ 10~G) not only further suppresses linear polarization at low $\theta$, keeping it below $m_l \sim 0.05$ (5~G) or $\sim 0.02$ (10~G), but also gives rise to deviations from the previously smooth $m_l(\theta)$ profile. The latter hints that structure below the resolution of sampled $\theta$ has arisen, particularly at larger $\theta$. With a 5~G magnetic field, the secondary EVPA flips occur for $\tau_f$ as low as 4.5. However, for a 10~G magnetic field, the Faraday Rotation generated EVPA flip occurs for an optical depth as low as 1.7, preceding the appearance of the Van Vleck angle. Unlike the Van Vleck angle, which appears first for high $\theta$ before approaching its characteristic value, the Faraday Rotation generated flip occurs first at low $\theta$, and migrates to higher $\theta$ with increasing $\tau_f$.

The failure of full solutions with $n_e = 1.57e5$ cm$^{-3}$ to converge for higher optical depths to the desired precision may be improved by increasing the resolution of the line of sight optical depth, $\Delta \tau$, though this would also cause a corresponding increase in compuation time. For the maximum Faraday Rotation case analyzed here, there are a total of six EVPA flips along the line of sight at $\theta = 2.5^{\circ}$ with an optical depth of 10. This leaves an average of only $\sim 16.8$ bins between successive EVPA flips along a single ray's path, which may result in degraded accuracy in the line of sight integration and the resulting full inversion solution. 

\section{Tests against previous work}

We consider here a number of comparisons between results of
the current work and those of earlier authors, from the
analytical results of GKK to the recent analysis by
\citet{2019A&A...628A..14L}, hereafter LV19.

\subsection{Comparison with GKK and Watson et al. models}

\begin{figure}[t!]
\begin{center}
\includegraphics[width=0.6\linewidth]{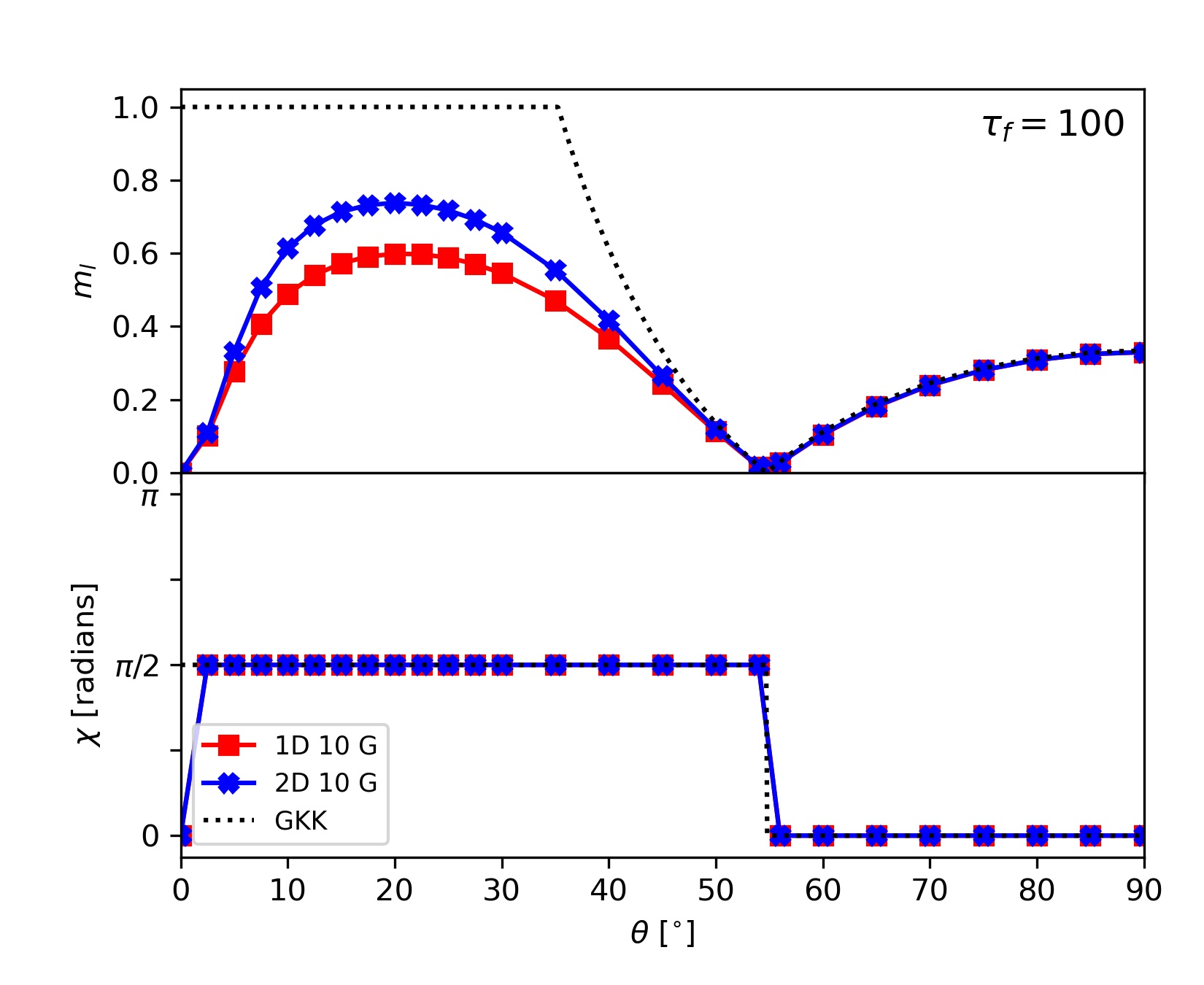}
\caption{Our linear polarization model results for a depth of $\tau_f=100$, plotted as a function of angle between the magnetic field and propagation directions, compared with the analytical expression from GKK (dotted black line). Results are shown from models with one- and two-directional propagation and no Faraday rotation, with a 10~G magnetic field. At $\tau_f=100$, there is no visible difference between different magnetic field strengths for a given number of propagating rays.}\label{fig:comp_vs_GKK}
\end{center}
\end{figure}

The original benchmark for our code, PRISM, is that it should be able to reproduce the results for both linear and circular polarization in \cite{2001ApJ...558L..55W}, even though the methods of analysis are not quite the same, and the numerical implementation is different. We additionally compare our results with the predictions of GKK, but we do not expect our code to achieve the GKK limits.

As a demonstration, we show in Figure~\ref{fig:comp_vs_GKK} the linear polarization fractions of two of our models, compared with the GKK prediction (in the appropriate limit of Doppler width $\gg$ Zeeman splitting $\gg$ stimulated emission rate, essentially GKK case 2a) as a function of the angle between the directions of propagation and the magnetic field. We note that the maser depth used ($\tau_f=100$) implies a level of saturation that exceeds the capacity of our model, but we use this high value to demonstrate the very slow convergence towards the GKK limit of 100\% linear polarization for angles significantly less than the Van Vleck value (see also Figure \ref{fig:mlevpa_v_theta_nofr_k01}). At angles greater than this value, the predictions of GKK and our model agree almost perfectly.

We do note that, as seen in Figure \ref{fig:comp_vs_GKK}, the fractional linear polarization at lower $\theta$ is higher when integrating along two bi-directional rays than with the single ray model. In the case of a
fully 3-dimensional model, saturation is enforced through an
angle-dependent function analogous to a solid-angle averaged
intensity. Early results\footnote{http://www.jb.man.ac.uk/~setoka/SEtoka\_EWASS2019\_ePoster.pdf} are not complete enough to show whether the trend of increasing $m_l$ with the number of rays continues at values of $\theta$ smaller than the Van Vleck angle.

\begin{deluxetable}{lcc}
    \caption{Saturation Conversion\label{tbl:conversions_for_watson}}
    \tablehead{
        \colhead{$i$} & \colhead{$\tau_f$ (1D)} & \colhead{$\tau_f$ (2D)}
    }
\startdata
        $10^{-2}$ & 3.5 & 3.5 \\
        $10^{-1}$ & 4.3 & 4.4 \\
        $1$ & 6.1 & 7.4 \\
        $3$ & 9.0 & 13 \\
        $10$ & 19 & 32 \\
        $10^2$ & $>100$ & $>100$ \\
        $10^3$ & $>100$ & $>100$ \\
        $10^4$ & $>100$ & $>100$  
    \enddata
\end{deluxetable}

We have computed unidirectional and bidirectional solutions to compare directly against the linear and circular polarization fractions presented in \cite{2001ApJ...558L..55W}. The saturation intensity used for normalization in that work, $I_{s,WW}$, and the saturation intensity defined here, $I_{sat}$, are equivalent, though the former is a specific intensity per frequency, while the latter is in specific intensity per angular frequency. Therefore, the dimensionless Stokes parameters presented here are directly comparable to the dimensionless Stokes parameters in \citet{2001ApJ...558L..55W}. The unpolarized seed radiation used in the calculations presented here, $i_0=10^{-8.2}$, is between the two values of $10^{-5}$ and $10^{-9}$ used as seed radiation in that work. In addition, they use the unitless Stokes I as a proxy for maser saturation. Table \ref{tbl:conversions_for_watson} outlines the conversion between unitless Stokes $i$ and the total optical depth required to reach that Stokes $i$ at line center for uni- and bi-directional integration.

\begin{figure}[t!]
\begin{center}
\includegraphics[width=0.92\linewidth]{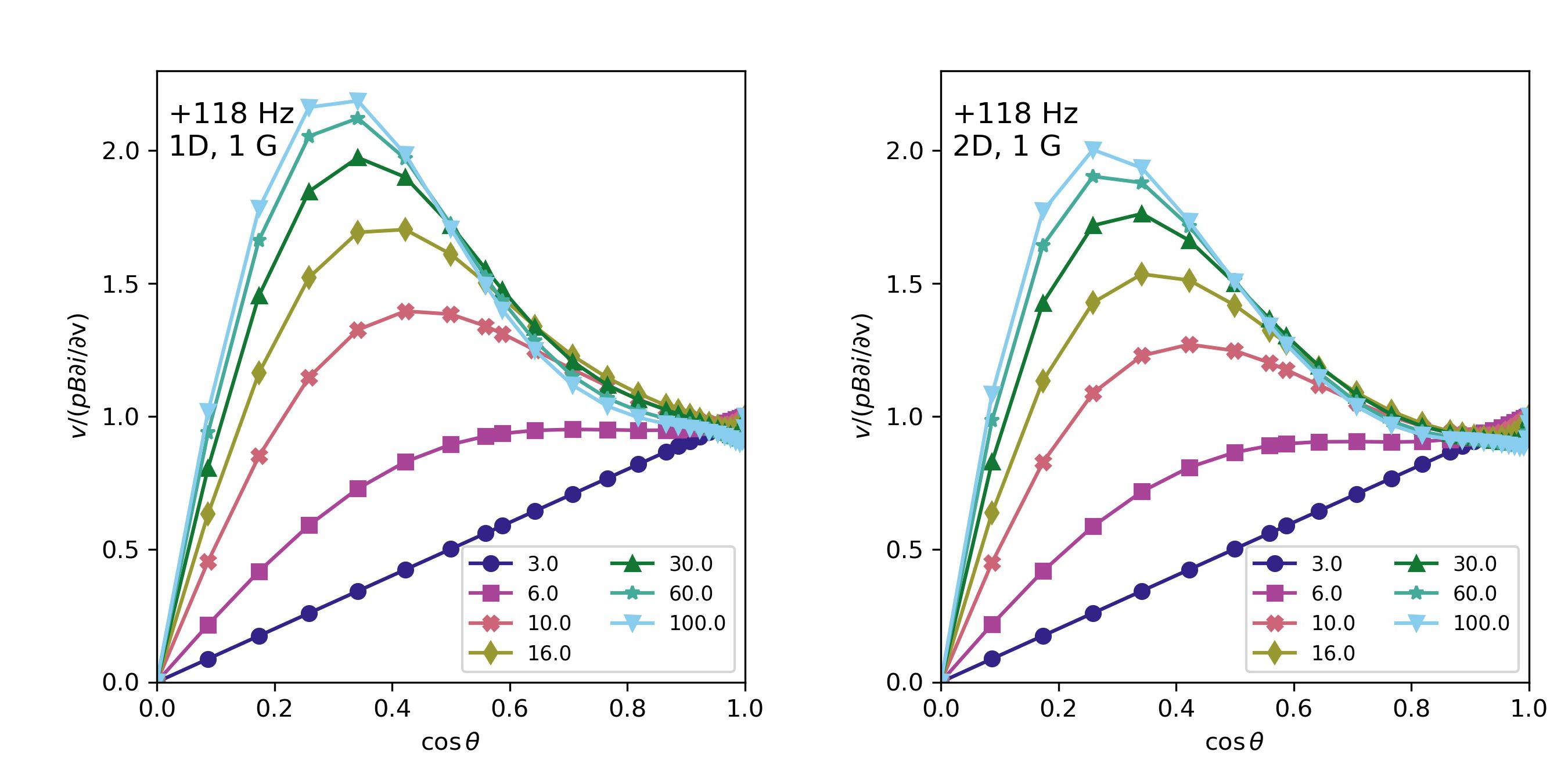}
\caption{Unitless Stokes $v$ normalized by the partial derivative of the unitless Stokes $i$ with respect to velocity and the Zeeman splitting in velocity space, $\Delta v_z = p B$, for comparison with \cite{2001ApJ...558L..55W} Figure 1. Values shown are computed at cloud end for a uni-directional (\textit{left}) or bi-directional (\textit{right}) cloud at a one bin offset from line center, with the partial derivative estimated via finite differencing using the two velocity bins on either side. The total optical depth of the cloud, $\tau_f$, for each curve is denoted in the legend.}\label{fig:watson_comp_v}
\end{center}
\end{figure}

For comparison with Figure 1 of \cite{2001ApJ...558L..55W}, the unitless Stokes $v$ normalized by the partial derivative of the unitless Stokes $i$ with respect to velocity and the Zeeman splitting in velocity space, $\Delta v_z = p B$, is plotted in Figure \ref{fig:watson_comp_v} for uni- and bi-directional masers. While the figures plotted are for 1~G magnetic fields, the normalization by the Zeeman splitting in velocity space removes any variation with $B$. Data plotted is calculated at a single bin offset ($\sim +118$ Hz) from line center, the normalization by the partial derivative removes any significant variation with frequency out to $\sim 256$ bins ($\sim + 30$ kHz).

As seen in Figure \ref{fig:watson_comp_v}, the framework presented here reproduces the dependence of circular polarization on $\theta$ from \cite{2001ApJ...558L..55W}. At low $\tau_f \lesssim 3$, $v / (p B \partial i / \partial v ) = \cos \theta$ as expected for their equivalent saturation of $I_{WW} = 10^{-2}$, with the dependence becoming peaked as optical depth or saturation increases. While the seed radiation used here is intermediate to the two values used in \cite{2001ApJ...558L..55W}, they note that the magnitude of the seed radiation only affects this dependence for $I_{WW} \gtrsim 10^2$, at which point the peaks with the larger seed radiation of $I_{0,WW} = 10^{-5}$ become $\sim 15 - 20 \%$ smaller than for the smaller seed radiation of $I_{0,WW} = 10^{-9}$. The intermediate equivalent seed radiation used in the solutions presented here yields a peak $v / (p B \partial i / \partial v ) \sim 1.96$ for a uni-directional integration with $\tau_f = 30$ and $\sim 1.85$ for a bi-directional integration with $\tau_f = 45$, both of which are within the range of peak values produced by the larger and smaller seed radiation from \cite{2001ApJ...558L..55W} for $I_{WW} = 10^2$.

\subsection{Comparison with Lankhaar and Vlemmings, 2019}

\begin{deluxetable}{l|cccc|cccc}
    \tablecolumns{9}
    \caption{Stimulated Emission Ranges\label{tbl:conversions_for_LV}}
    \tablehead{
        \colhead{$B$} & \multicolumn{4}{c}{$\log \left( R / g \Omega \right)$} & \multicolumn{4}{c}{$\log \left( R_0 / g \Omega \right)$} \\
        \colhead{} & \colhead{1D Min} & \colhead{1D Max} & \colhead{2D Min} & \colhead{2D Max} & \colhead{1D Min} & \colhead{1D Max} & \colhead{2D Min} & \colhead{2D Max}
    }
\startdata 
        1~G & -5.5 & 4.1 & -5.2 & 3.8 & -9.1 & 0.8 & -8.9 & 0.5 \\  2~G & -5.8 & 3.8 &   -  &  -  & -9.4 & 0.5 & - & - \\  
        5~G & -6.2 & 3.4 & -5.9 & 3.1 & -9.8 & 0.1 & -9.6 & -0.2 \\ 
        10~G & -6.5 & 3.1 & -6.2 & 2.7 & -10.1 & -0.2 & -9.9 & -0.5
    \enddata
\end{deluxetable}

In a work that draws heavily on earlier models in the
Watson et al. series (see above), LV19 have developed the very able
code {\sc champ} for the analysis of a wide range of
molecules and transition types. As such, it goes, in some
respects, considerably beyond the scope of the present work.
However, there are significant differences in the theoretical construction and numerical implementation that
should be discussed, and a successful agreement in the
principal results then strongly suggests that the underlying theory is correct, independently of the exact methods used.

Apart from the obvious improvements in LV19 relating to more general Zeeman patterns and hyperfine structure, and our inclusion of a counter-propagating ray and Faraday rotation, the main
differences between the LV19 analysis and that in the current work can be summarised as:
\begin{enumerate}
    \item In LV19, the time-dependence of off-diagonal elements of the DM is integrated out, and a subsequent steady-state approximation applied to the level populations to reduce the system to an algebraic (matrix) equation. In the present work, we follow rather the method of \cite{2009MNRAS.399.1495D} in Fourier-transforming the equations to the frequency domain, eliminating the time-dependence of all DM elements in the process (see work leading to our equations \ref{eq:dmoff1} - \ref{eq:diag1pi}) for a finite sampling time of the order of the reciprocal channel width.
    \item The populations are recovered from a matrix inversion procedure in LV19; in the present work we make a formal solution of the coupled radiative transfer equations in order to algebraically eliminate the Stokes parameters, leaving a set of coupled algebraic equations in the inversions, our eq.(\ref{eq:sig_dimless}) and eq.(\ref{eq:pi_dimless}). The numerical solutions in the two works therefore rely on different sets of algorithms.
    \item The classical saturation approximation (described at the beginning of Appendix~\ref{a:appendix1}) is probably equivalent to the `sharply-peaked' approximation to the homogeneous lineshape function mentioned in LV19, and used in \cite{2001ApJ...558L..55W}. If this is so, the results presented in LV19 use a more general approximation, based on a Taylor expansion of the lorentzian profile, and {\sc champ} can therefore be used at higher (but not arbitrary) levels of saturation than the present work.
    \item Parameters used by the codes differ somewhat. LV19 scales maser saturation directly, using the base-10 logarithm of the ratio of the stimulated emission and Zeeman rates, rather than indirect adjustment as a result of the changing optical depth, $\tau_f$. We discuss below some problems related to the definition of the stimulated emission rate.
    \item The formalism presented here sets the type 2 off-diagonal elements of the DM to zero, unlike the results presented in LV19 from their method iii. See Section \ref{ss:rogue_dm} for a discussion of the affect of this approximation.
\end{enumerate}

Before introducing a formal definition of the stimulated emission rate, we note that historically it has been based on equations like eq.(44) of \citet{1990ApJ...354..660N} in which the Stokes parameters are dimensionally specific intensities, and outside this subsection we also refer to stimulated emission rates in this sense. In eq.(44) of \citet{1990ApJ...354..660N}, their $R$ is consistent with the stimulated emission rate as an Einstein B-coefficient multiplied by a line-center specific intensity (and trigonometric functions via the various dipole
products). This interpretation is backed by the expression for the low-intensity boundary condition on Stokes I near the beginning of Section~III of \citet{1990ApJ...354..660N}. We now define the stimulated emission rates 
used in the present work.

The stimulated emission rate per substate transition in each angular frequency bin, $R_n^{\pm,0}$, can be calculated for the solution presented here from the unitless Stokes solution at the end of the cloud:
\begin{equation}
    R_n^0 = 2 \Gamma \sin^2 \theta \left[ i_n - q_n \cos \left( 2 \phi \right) - u_n \sin \left( 2 \phi \right) \right] 
\label{eq:Rn0}
\end{equation}
\begin{equation}
    R_n^{\pm} = \Gamma \eta^{\pm} \left[ \left( 1 + \cos^2 \theta \right) i_{n \mp k} + \left( q_{n \mp k} \cos \left( 2 \phi \right) + u_{n \mp k} \sin \left( 2 \phi \right) \right) \sin^2 \theta \mp 2 v_{ n \mp k} \cos \theta \right] ,
\label{eq:Rnpm}
\end{equation}
noting that our $R_n$, and specifically the line-center $R_0$, are consistent 
with the traditional stimulated emission rate from
\citet{1990ApJ...354..660N} since the dimensionless Stokes parameters are scaled to the specific intensity,
$I_{sat}$ from eq.(\ref{eq:isat}). However, as we have full spectral information, we choose to write the stimulated emission rate calculated at line center as
\begin{equation}
    R_0 = R_0^- + R_0^0 + R_0^+
\end{equation}
and define a separate, total stimulated emission rate that encompasses the full spectral emission:
\begin{equation}
    R = \sum_{n} \left[ R_n^- + R_n^0 + R_n^+ \right].
\end{equation}

The stimulated emission rate in the $\varpi_n$ bin, $R_n^{\pm,0}$ may be viewed alternately as the rate, in s$^{-1}$, per angular frequency integrated over the width of the angular frequency bin: $R_n^{\pm,0} / \delta \varpi \times \Delta \varpi$. Then, summing the mean stimulated emission rate over the $n$ angular frequency bins gives a measure of the total stimulated emission rate represented by the full spectral linewidth.

We continue using $\Gamma = 5 \textrm{ s}^{-1}$ for consistency, though it is worth noting that the dimensionless solutions presented here with no Faraday Rotation are independent of $\Gamma$. The total Zeeman width is calculated as $g\Omega = 2k \Delta \varpi$. Calculating $R$ for each solution, we find that $\log \left( R / g \Omega \right)$ varies $\lesssim 1 \%$ as a function of $\theta$ alone, and may therefore be viewed as a proxy for $\tau_f$ within a given parameter set. The ranges of $\log\left(R/g\Omega\right)$ values calculated across all $(\theta,\tau)$ for a given magnetic field strength and number of rays are shown in Table \ref{tbl:conversions_for_LV}, as well as the values calculated at line center, $\log\left(R_0/g\Omega\right)$. In general, $\log\left(R/g\Omega\right)$ decreases with increasing magnetic field strength and is slightly lower for a bi-directional integration than for a uni-directional integration, but shows no significant variation with the inclusion of nonzero Faraday rotation. 

The ranges of values calculated at line center, $\log\left(R_0/g\Omega\right)$, follow the same trend, with typical values $\sim 3.3-3.7$ below $\log\left(R/g\Omega\right)$, though it is not a direct, uniform scaling. We discuss the relation between $R_0$ and $R$ further in Section \ref{ss:sigv}. For the purposes of this comparison with LV19, we will use the line center value, $R_0$, as they do, for a more direct comparison. However, we note that, while the non-linearities in the relation between $R$ and $R_0$ do cause minor adjustments in the precise form of the results presented as a function of $R_0$ and $R$ here, the general trends are unchanged.

\begin{figure}[t!]
\begin{center}
\includegraphics[width=0.94\textwidth]{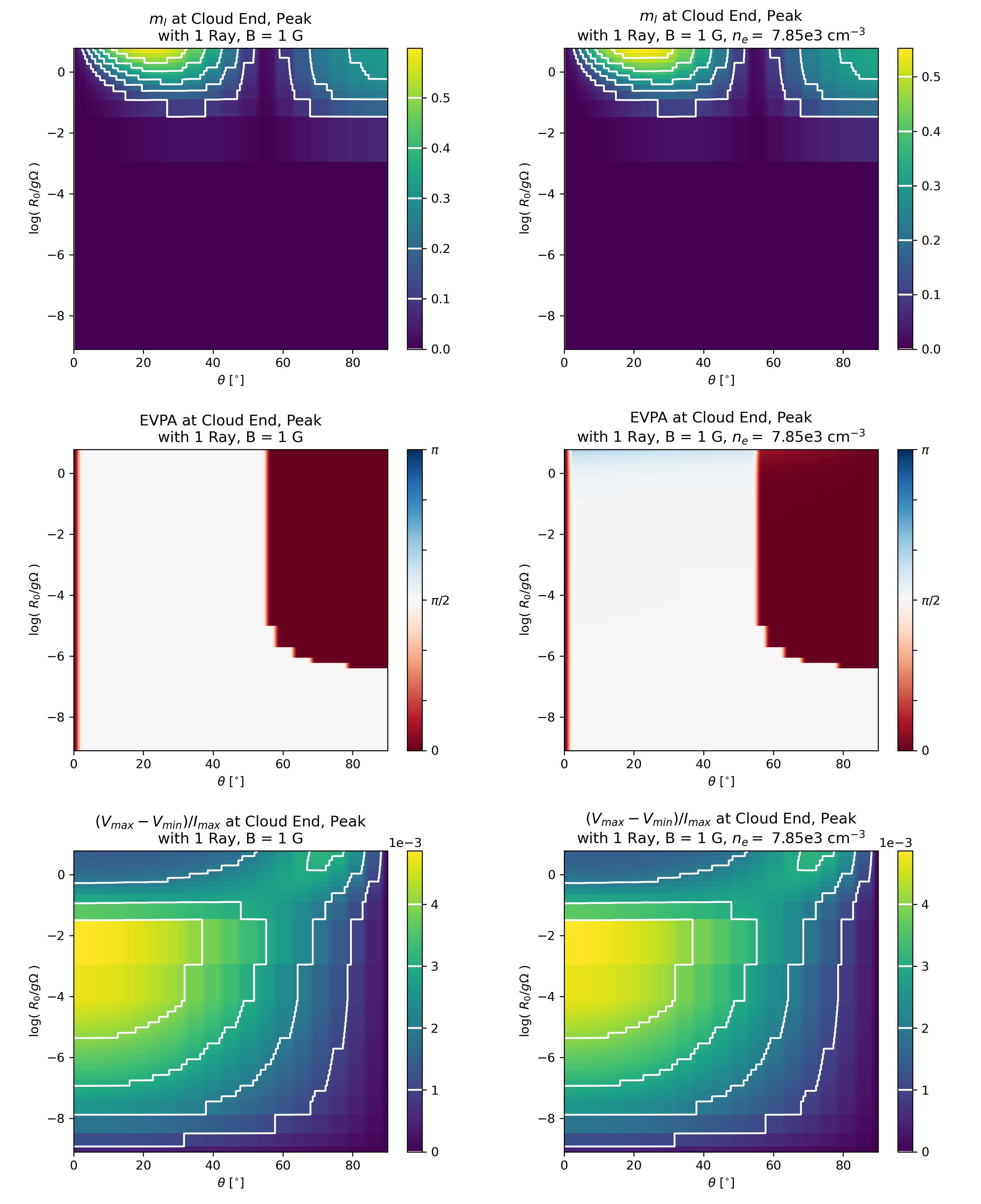}
\caption{(\textit{Top}) Line center $m_l$, (\textit{middle}) line center EVPA, and (\textit{bottom}) range in Stokes $v$ across $\varpi$ normalized by the line center Stokes $i$. Values are shown at cloud end and shown as a function of $\theta$ and calculated $\log \left( R_0 / g \Omega \right)$ for the 1-directional solution set with $B = $ 1~G with (\textit{left}) no Faraday rotation, and (\textit{right}) Faraday rotation with $n_e = 7.85e3 \textrm{ cm}^{-3}$. }\label{fig:LandV_1d_k1}
\end{center}
\end{figure}

Figure \ref{fig:LandV_1d_k1} shows the line center $m_l$ and EVPA for the 1-directional, $B=$ 1~G solutions with no Faraday rotation and Faraday rotation with $n_e = 7.85e3 \textrm{ cm}^{-3}$, as well as their respective ranges in Stokes $v$ across $\varpi$ normalized by line center stokes $i$ (referred to as $p_C$ in LV19) for comparison. 

While the ranges of $\log \left( R_0 / g \Omega \right)$ covered by our solutions differ from those of LV19, the most notable difference in linear polarization between the solutions presented here and those of LV19 in the overlapping region (-3 to 0.8) regards the onset and/or disappearance of the EVPA flip with increasing maser saturation. The 1~G SiO solutions of LV19 have a $\pi/2$ EVPA rotation at the Van Vleck angle only occurring for low $\log \left( R_0 / g \Omega \right) \sim$ -2 to -3, which smooths out and disappears with increasing $\log \left( R_0 / g \Omega \right)$. Conversely, the solutions presented in this work show no smoothing or disappearance of the EVPA flip with increasing saturation between $\log \left( R_0 / g \Omega \right) \sim -2$ up to our maximum 0.8. At lower saturation than that covered by LV19, our solutions also show the $\pi/2$ instantaneous EVPA flip appearing for large $\theta$ and approaching the Van Vleck angle as $\log \left( R_0 / g \Omega \right)$ increases.

\begin{figure}[t!]
\begin{center}
\includegraphics[width=0.97\textwidth]{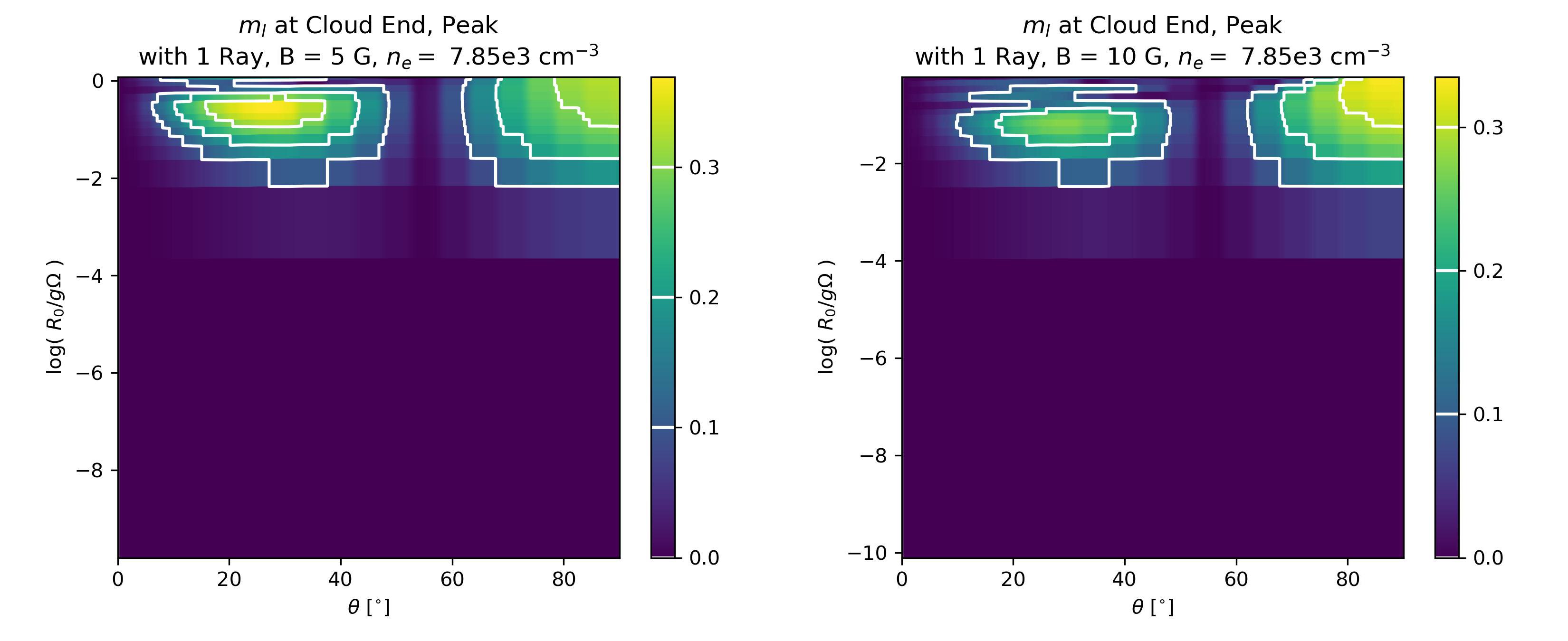}
\caption{Line center $m_l$ at cloud end as a function of $\theta$ and calculated $\log \left( R_0 / g \Omega \right)$ for the 1-directional solution set with $n_e = 7.85e3 \textrm{ cm}^{-3}$ for (\textit{left}) $B =$ 5~G and (\textit{right}) $B =$ 10~G. }\label{fig:LandV_ml_1d_fcA}
\end{center}
\end{figure}

The difference in the appearance and behavior of the EVPA rotation also causes a marked difference in the $m_l$ profile, with the $m_l=0$ minimum at the Van Vleck angle persisting in our solutions to the largest computed values of $\log \left( R_0 / g \Omega \right)$, once it has set in. The LV19 solutions also show a distinct peak and subsequent decrease in $m_l$ for increasing $\log \left( R_0 / g \Omega \right)$, with the peak occurring at $\log \left( R_0 / g \Omega \right) \sim 0$ as in the 1~G SiO solution.

While the solutions presented here only go up to $\log \left( R_0 / g \Omega \right) \sim 0.8$, we see no indication of a decrease in $m_l$ with increasing saturation above $\log \left( R_0 / g \Omega \right) \sim 0$ in our solutions with no, or weak, Faraday rotation. However, this behavior is visible in our solutions with stronger Faraday rotation, but it occurs at lower values of $\log\left( R_0 / g \Omega \right)$ (eg. Figure \ref{fig:LandV_ml_1d_fcA}). Although not present in the calculated solutions for $n_e = 7.85e3 \textrm{ cm}^{-3}$ with a low $B = \textrm{1~G}$, $m_l$ does begin decreasing at the highest levels of $\log \left( R_0 / g \Omega \right)$ achieved here, as the Faraday rotation begins suppressing linear polarization. The suppression of $m_l$ at high $\log \left( R_0 / g \Omega \right)$ increases significantly for $n_e = 1.57e5 \textrm{ cm}^{-3}$ particularly for the peak at lower $\theta$. Notably, the solutions from LV19 method (iii) do allow for Stokes QU conversion by allowing for non-zero $E(\varpi)$, much like our inclusion of non-zero Faraday rotation with $\gamma_{qu}$. Therefore, the comparison between the solutions derived here and those of LV19 may be most appropriately applied to our solutions with nonzero Faraday rotation.

In addition to terms linking Stokes $U$ to $I$ and $Q$, the condition 
$R_0 \gtrsim g\Omega$ also introduces terms into the maser propagation equations that couple
$U$ and $Q$ to $V$. These all arise from a shift of the `good' symmetry axis for quantization away
from $\vect{B}$ and towards the propagation axis as $R_0/(g\Omega)$ increases. If the magnetic
field is still used as the quantization axis when $R_0 \gtrsim g\Omega$ then these compensatory terms
enter through the type~2 off-diagonal elements of the DM, and such terms are included in LV19 and
in \citet{1994ApJ...423..394N}.
However, we also note that, while the line center $R_0$ values only extend up to $\log \left( R_0/g\Omega \right) \sim -0.5$ to $+0.8$ in the solutions presented here, the stimulated emission calculated using the full line breadth, $R$, are $\sim 3.3$ to $3.7$ times the order of magnitude of $R_0$. 

The circular polarization metric used by LV19, $p_C = ( V_{max} - V_{min} ) / I_{max}$, is likewise shown in Figure \ref{fig:LandV_1d_k1}. The isotropic LV19 solutions for SiO in a 1~G field have $p_C$ reaching a peak of $p_C \gtrsim 0.25$ around $\theta \sim 20^{\circ}$ at $\log \left( R_0/g \Omega \right) \sim 0.8$. Starting at saturation $\log \left( R_0/g \Omega \right) \sim -0.5$, their $p_C$ has reached $\sim 0.5$ for $\theta \sim 10^{\circ}$ to $70^{\circ}$. In our solutions, with and without Faraday rotation, $p_C$ shows a marked peak and decrease at low $\theta$ with increasing $\log \left( R_0 / g \Omega \right)$, with the peak occurring around $\log \left( R_0/g\Omega \right) \sim -3$ to $-2$ for our 1~G solution. However, our solutions have no decrease in $p_C$ for small $\theta$ as $\theta \rightarrow 0^{\circ}$. Our solutions also show a secondary, lower amplitude peak in $p_C$, peaking for $\theta \sim 70^\circ$ at the highest $\log \left( R_0 / g \Omega \right)$. 

Notably, the values of $p_C$ presented here for a 1~G magnetic field are below the levels of the contours in LV19. Despite measuring the full peak-to-valley range of Stokes v, $p_C$'s normalization by the line center $i_0$ rather than the value of Stokes $i$ at the same frequency at which Stokes $v$ is also being measured results in $p_C$ being significantly lower than the $m_c$ presented previously in this paper. However, our 1~G solutions still lack the higher values of LV19 for $\log \left( R_0/g\Omega \right) \gtrsim -0.5 $ for most $\theta$. In addition, our B$=$10~G solutions show the same behavior in $p_C$ as the 1~G solutions presented in Figure \ref{fig:LandV_1d_k1} with $p_C$ values extending up to $\sim 0.05$. While the saturation regime for those only extends up to $\log \left( R_0 / g \Omega \right) \lesssim -0.2$, the secondary peak in our solutions that starts forming at high $R_0 / g \Omega $ mentioned above, remains limited to high values of $\theta$, and, at $p_C \gtrsim 0.03$, still falls shy of reaching the level of the LV19 contours for the same magnetic field.

As discussed below in section \ref{ss:rogue_dm}, the assumption by the present work that type~2 off-diagonal DM elements are negligible compared to those that represent the electric dipole-allowed transitions may explain some of these discrepancies at the higher values of $\log(R_0/g\Omega)$.

\section{Discussion}
\label{s:disco}

We have demonstrated that our maser polarization code PRISM behaves as expected as the Zeeman splitting is varied from very small values, of order similar to the width of the homogeneous response profile of the molecules, to values that substantially exceed the inhomogeneous (Doppler) width. In the latter case, the solutions correspond to the circularly-polarized Zeeman doublet when the field and propagation directions are parallel, and to a linearly-polarized triplet when these directions are perpendicular. We have compared results with the earlier work of GKK (finding agreement in the expected range of angles), with the work of Watson's group, where we have demonstrated agreement for uni- and bi-directional masers at stimulated emission rates up to approximately $R_0/(g\Omega)=10^4$, where $R_0$ implies $R_n$ at line center, and finally with the recent predictions of the {\sc champ} code, noting that we have restricted our analysis to lower levels of saturation than those in \citet{2019A&A...628A..14L}. Overall, we consider the degree of agreement between the various models to be very good, considering the differences in the analysis strategies and numerical algorithms. We consider specific areas of further discussion below.

\subsection{Consequences of approximations}
\label{ss:conseq}

The main approximations that limit the degree of saturation accessible to our code are those that form part of the `classical saturation' set (see Appendix~\ref{a:appendix1}). We have limited our examples to modest levels of saturation on account of these simplifications. The first of these, no correlation between different Fourier components of the radiation field, is not problematic unless the input radiation to the maser is itself coherent. However, the second approximation must be lifted as a prerequisite for a fully semi-classical response. This requires working with the unsimplified eq.(\ref{eq:diag1sig}) and eq.(\ref{eq:diag1pi}) as the descriptions of the inversions, noting that, in this pair of equations, the maser field is represented by electric field amplitudes. An expression in terms of the Stokes parameters is only possible if the third assumption (gaussian statistics) is maintained, but departures from gaussian behaviour are to be expected, starting at line centre, and moving towards the wings at larger signal strengths \citep{2009MNRAS.396.2319D,2009MNRAS.399.1495D}. Replacement of the Dirac $\delta$-function approximation with full lorentzian homogeneous lineshape functions is therefore of limited value in increasing accessible levels of saturation without making more significant changes to the code to incorporate fully semi-classical saturation.
Equations (\ref{eq:diag1sig}) and (\ref{eq:diag1pi}) are not completely intractable. One approach, that
follows the style of solution in the present work, is to make a formal solution of the 
frequency-domain electric field complex amplitudes, eliminating these from eq.(\ref{eq:dmoff1})-eq.(\ref{eq:diag1pi}) in favor of off-diagonal DM elements. 
Then, eq.(\ref{eq:diag1sig}) and eq.(\ref{eq:diag1pi}) are used to 
eliminate the inversions from eq.(\ref{eq:dmoff1}) and eq.(\ref{eq:dmoffpi1}), leaving a set of non-linear algebraic equations in the
{\em off-diagonal} DM elements. Another possibility is to use the method derived in
\citet{2021MNRAS.507.4464W}, which is a generalisation of the Fourier-based method developed by
\citet{1978PhRvA..17..701M}.

\subsection{Accessible range of saturation}

The limit placed on the degree of saturation in \citep{2001ApJ...558L..55W} is that the stimulated emission rate must be considerably less than the Zeeman splitting ($R_0 \ll g\Omega$). This limit is mentioned in the text following their eq.(12), and is applied because they use classical rate equations rather than a semi-classical (DM) representation of the molecular response. For our models with $\Gamma = 5$\, s$^{-1}$ and magnetic fields of 1-10~G, we have $R_0/(g\Omega)$ in the approximate
respective range 1/3-1/30 at an optical depth of $\tau_f = 30$. Our use of an upper limit optical depth of 30 therefore conforms to the $R_0 \ll g\Omega$ for the larger magnetic fields, but is becoming marginal for $B=$ 1~G. At an optical depth of 100, the approximation fails, even for the highest field used.

\citet{1990ApJ...354..660N} extend the range of accessible saturation by adopting a semi-classical molecular response. As the present work also uses a DM, we discuss here whether we may also extend our saturation limit beyond the $R_0 \ll g\Omega$ constraint. There are two possible reasons why we should not do this: the first is the presence of off-diagonal DM elements within a single $J$-state of the molecule. However, \citep{1990ApJ...354..660N} also ignore these, so we defer discussion of them to
Section~\ref{ss:rogue_dm} below. The remaining reason that might limit our degree of saturation is the use of classical reductions (see Section~\ref{ss:conseq}), particularly the replacement of lorentzian homogeneous profiles by $\delta$-functions. For a brief understanding of the consequences of this, we would always want to resolve the Zeeman splitting, resulting in maximum channel widths of order $g\Omega$ in frequency. Lorentzian profiles power broaden according to $\Gamma' = \Gamma \sqrt{1 + I/I_{sat}}$, which, for strong saturation, may be inverted to yield $I/I_{sat} \simeq (\Gamma' /\Gamma)^2$, where $\Gamma$ is the original width parameter. Setting $\Gamma' = g\Omega$, and $I/I_{sat} \simeq R_0/\Gamma$, we
then find we require $R_0 < (g\Omega)^2/\Gamma$, exactly the criterion adopted for the upper limit of magnetically induced polarization in GKK. We also note that \citet{1990ApJ...354..660N} continue to use the magnetic field as a quantization axis in this regime, without moving to the ray-based quantization of the GKK Case 3, where $R_0 \gg g\Omega$. Adopting the $R_0 < (g\Omega)^2/\Gamma$ limit {\it might} allow us to access saturation levels as high as perhaps 10\% of
$(1500/5)^2 = 90,000$, or 9000 for a 1\,G field. However, a more accurate analysis suggests greater caution is required.

A power-broadened homogeneous lineshape is described, for example, in \citet{2001OptCo.199..117V}, and a unity-normalized version in offset frequency $\varpi$ looks like,
\begin{equation}
    L(\varpi) = \frac{2[1+j]^{1/2}}
                 {\pi \Gamma [1 + j + (\varpi /\Gamma)^2]},
                 \label{eq:power_b}
\end{equation}
where $j=J/I_{sat}$ for mean intensity $J$ and saturation intensity $I_{sat}$, and $\Gamma$, as before, is the unsaturated homogeneous width. Suppose that we wish to limit semi-classical effects to population pulsations with magnitudes smaller than 10\% of the total inversion: we can obtain the corresponding half-width of the power-broadened function by integrating $L(\varpi)$ from eq.(\ref{eq:power_b}) over the limits zero to the desired half width, $\varpi_{\%}$, and equating the result to 0.95. The result of the integration is
\begin{equation}
 \varpi_{10\%}/\Gamma = 12.71 [1+j]^{1/2} .
 \label{eq:widpulse}
\end{equation}
The width is rather large, owing to the relatively broad wings of the lorentzian function compared to a gaussian.
We now set this width equal to the actual Fourier channel widths used. For the SiO parameters applied to the models in Figure~\ref{fig:mlevpa_v_theta_nofr_k01}, the channel width is $\delta \omega = 740.5$\,s$^{-1}$, so using the value of 5\,s$^{-1}$ for $\Gamma$ introduced above, we arrive at $j=I/I_{sat}=135$. This figure is not much less restrictive than that imposed by $R_0 \ll g\Omega$, but probably means that $j=100$,
the largest degree of saturation 
used in this work is acceptable, but higher values should be modeled only if the simplifications described in Section~\ref{ss:conseq} are lifted. Our value of $j$ derived above is also consistent with the figure of 100 stated in Section~5.1.1 of LV19.

\subsection{Off-Diagonal DM elements}
\label{ss:rogue_dm}

Off-diagonal elements of the DM may be divided into two types: those (type 2) that represent the coherence
between levels that would be degenerate in the absence of an applied magnetic field, and the remainder
(type 1) between levels that would not.
The center frequency of a type~2 transition is therefore comparable to the spread of a single Zeeman group, or a few Doppler widths at most, whilst the center frequency of a type~1 transition is comparable to the rest frequency of the unsplit transition, $J=1-0$ in the present work. Transitions of the two types behave differently when the rotating wave approximation is applied. The off-diagonal DM element of a type~2 transition satisfies a differential equation in the time domain that has no direct coupling to an inversion: it can develop only from itself and from other off-diagonal DM elements. In the $J=1-0$ system considered in the present work, there are three possible type~2 DM elements, each of which is coupled directly to a pair of off-diagonal DM elements representing electric dipole-allowed transitions.

We note that in the present work the type~2 off-diagonal elements are set to zero. However, \citet{2019A&A...628A..14L} do not make this assumption, and include the type~2 elements, as do \citet{1990ApJ...354..660N}, where they are assumed to be constant when solving for the DM elements
(their eq.(10)).
We can estimate the consequences of ignoring the type~2 DM elements by including the correction due to these terms in a Fourier component of a type~1 element. With these corrections, a reduced version of eq.(\ref{eq:dmoff1}) for $s_n^+$ with $\cos \theta = \cos \phi =1$, $m=n$, and considering one hand of polarization only, takes the form
\begin{align}
s_n^+ & = \frac{\sqrt{2} \pi \hat{d}^+ \amp{E}_n \tilde{L}_n^+}
                 {\hbar}
                 \left\{
                 \Delta_0^+ \left[
                 1 - \frac{\pi^2 \tilde{L}_n^+}{2\hbar^2}
                 \sum_{k=-\infty}^\infty 
                 \left(
                 \tilde{L}^{*,[0,-1]}_{k-n} (\hat{d}_k^0 \amp{E}_k)^2 
                 +
                 \tilde{L}^{*,[1,-1]}_{k-n}
                 (\hat{d}_k^- \amp{E}_k)^2
                 \right)
                 \right] \right. \nonumber \\
                & \left.
     -\frac{\pi^2}{2\hbar^2} \sum_{k=-\infty}^\infty \left(
    \Delta_0^0 \tilde{L}_{k}^{0,*}
               \tilde{L}_{k-n}^{*,[0,-1]}
               (\hat{d}^0 \amp{E}_k)^2
   +\Delta_0^- \tilde{L}_{k}^{-,*}
               \tilde{L}_{k-n}^{*,[1,-1]}
               (\hat{d}^- \amp{E}_k)^2
                             \right)
    \right\} ,            
\label{eq:swith_type2}
\end{align}
where complex lorentzian functions corresponding to type~2 transitions have their $m_J$ quantum-number pairs shown in square brackets in the superscript. Equation~\ref{eq:swith_type2} demonstrates that the terms within the sums over $k$ that result from the inclusion of type~2 off-diagonal DM elements are of order $(\hat{d} \amp{E})^2$ with respect to 1, and therefore are likely to become significant under strongly saturating conditions. Further, the type~2 contributions are modified by the product of pairs of complex lorentzian functions so their effect will increase as these functions power-broaden. The effect of the type~2 terms is deleterious to the coherence of the $\sigma+$ transition, both reducing the coupling to its own inversion (top line in eq.(\ref{eq:swith_type2})), and by introducing population mixing via terms involving inversions in the other two type~1 (dipole-allowed) transitions (lower line in eq.(\ref{eq:swith_type2})). Overall then, the type~2 off diagonal elements will reduce the polarization at very high degrees of saturation, even if they are initially set to zero, and are an additional reason why our current model will become unreliable for stimulated emission rates significantly higher than $g\Omega$.

\subsection{The significance of nonzero Stokes V}
\label{ss:sigv}

\begin{figure}[t!]
\begin{center}
\includegraphics[width=0.97\textwidth]{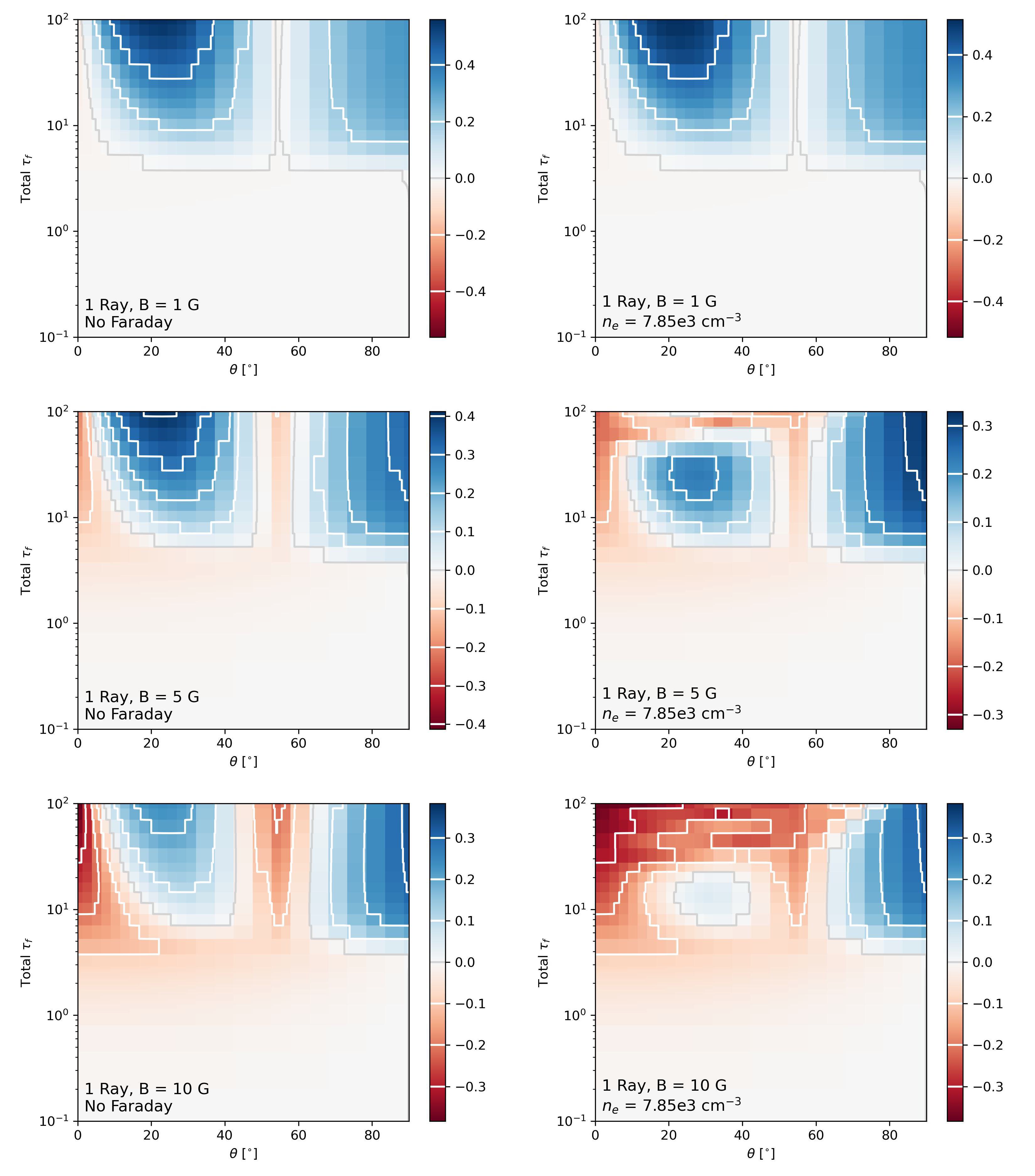}
\caption{Difference between the peak $m_l(\varpi)$ and maximum $m_c(\varpi)$ at cloud end, shown as a function of $\theta$ and total optical depth, $\tau_f$, for uni-directional solutions. Magnetic field strength and $n_e$, if Faraday rotation was utilized, are labelled in each figure. In each figure, the gray contour indicates $m_l(\varpi=0) = max\left( m_c (\varpi) \right)$. }\label{fig:mlmc}
\end{center}
\end{figure}

Many previous publications related to maser polarization for small Zeeman splitting (much smaller than the Doppler width) ignore circular polarization entirely, either for simplicity, or citing the antisymmetric profile of Stokes~V. GKK do discuss off-resonance propagation, but their strong result in Case 2 is of zero circular polarization at line-center. Another major work that considers only line-center amplification is \citet{1990ApJ...354..649D}, and this results in linear polarization only.
The formalism presented here includes the propagation of Stokes $V$, as well as the retention of the full Stokes profile as a function of angular frequency, $\varpi$. This allows a comparison of not only the relative intensities of the linearly- and circularly-polarized emission, but also a comparison of the maximum values of $m_l (\varpi)$ and $m_c (\varpi)$ achieved, as the anti-symmetric profile of Stokes $V(\varpi)$ necessitates that its peak occurs away from the peak in Stokes $I(\varpi)$ at line center. 

To provide a sense of the relative amplitudes of $m_l(\varpi)$ and $m_c(\varpi)$ for the derived solutions, we plot $m_l(\varpi=0) - max \left( m_c (\varpi) \right)$ as a function of $\theta$ and total optical depth, $\tau_f$, for the uni-directional solutions for three magnetic field strengths, with and without Faraday rotation (Figure~\ref{fig:mlmc}). At high $\tau_f$, $max\left( m_c (\varpi) \right)$ is still larger than $m_l (\varpi=0)$ around $\theta = 0^\circ$ and the Van Vleck Angle, where $m_l \rightarrow 0$. For solutions with stronger Faraday rotation, $max\left( m_c (\varpi) \right)$ surpasses $m_l (\varpi=0)$ for large $\tau_f$ as the Faraday rotation suppresses $m_l(\varpi)$.

Notably, in all cases with $\theta \neq 90^\circ$, the maximum $m_c(\varpi)$ is larger than the peak $m_l(\varpi=0)$ for $\tau_f \lesssim 3$. This trend does not appear to be driven by an amplification of $m_c$ from the decreased Stokes $i$ away from line center; a similar characteristic behavior is seen when comparing the peak $\sqrt{q^2 + u^2}$ and maximum $v(\varpi)$. This behavior implies that, particularly at low saturation ($\tau_f \lesssim 3$), circular polarization is not negligible compared to linear polarization.

\begin{figure}[t!]
\begin{center}
\includegraphics[width=0.97\textwidth]{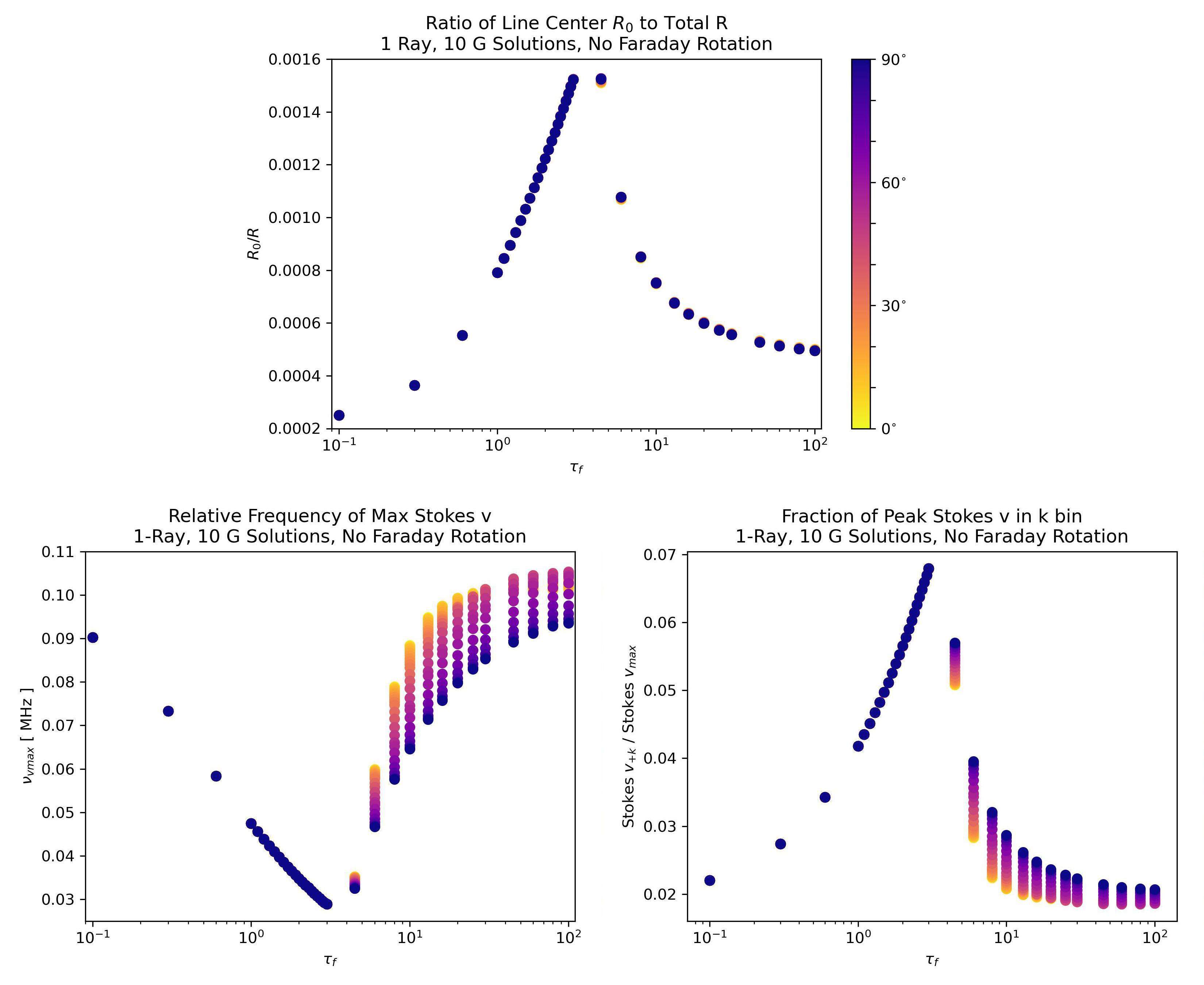}
\caption{Three figures showing \textit{(top)} the ratio between the line center and summed stimulated emission rates, $R_0/R$, \textit{(lower left)} the frequency offset from line center at which the peak in Stokes $v$ occurs, and \textit{(lower right)} the ratio between the Stokes $v$ in the bin $k$ from line center and the peak Stokes $v$ value, all as a function of the total optical depth, $\tau_f$. All figures shown are for the uni-directional solution with $B=10$~G and no Faraday Rotation, with the color indicating the angle $\theta$ of the plotted solution. }\label{fig:R0R_comp}
\end{center}
\end{figure}

Another benefit of accounting for Stokes $v$ away from the line center is that it provides insight into the relation between the stimulated emission rate calculated at line center, $R_0$, and the stimulated emission rate summed over the full line profile, $R$. As shown in Figure \ref{fig:R0R_comp}, the relation between $R_0$ and $R$ as a function of $\tau_f$ is significantly non-linear across the range of solutions presented here, increasing by a factor of $\sim 5$ from $\tau_f = 0.1$ to $3.0$. 

While this would have relatively little effect on the functional form of the results plotted in, say, Figure \ref{fig:LandV_1d_k1} and \ref{fig:LandV_ml_1d_fcA} if they were plotted using $R$ instead of $R_0$ (simply distorting some of the vertical scaling), it does suggest that caution may be warranted when using a stimulated emission rate calculated at line center, $R_0$ to infer the strength of the full stimulated emission rate, $R$, even if $R_0$ includes nonzero Stokes $v$ at bins $\pm k$ from the line center.

As can be seen in eq.\ref{eq:Rn0}, Stokes $v$ is included in the $R_0$ calculation in the form of $v_{\mp k}$, despite the value of Stokes $v$ at line center being zero. Then, the ability for $R_0$ to accurately reflect the full $R$ is limited by the ability of Stokes $v_{\mp k}$ to trace the peak strength of Stokes $v(\varpi)$. However, the profile of Stokes $v(\varpi)$ can vary not only in amplitude, but also in the offset frequency from line center at which it peaks. Figure \ref{fig:R0R_comp} also shows how the frequency at which Stokes $v$ peaks and the ratio between Stokes $v_{+k} / v_{max}$ varies with $\tau_f$. As $\tau_f$ increases from 0.1 to $\sim 3$, the Stokes $v(\varpi)$ profile becomes narrower, and the fraction of the peak $v_{max}$ that occurs in the bin $k$ from line center increases. The Stokes $v_{\mp k}$ term in $R_0$ contains a larger fraction of the total Stokes $v$, and, as a result, $R_0$ increases with respect to $R$. Then, as $\tau_f$ continues increasing beyond $\sim 4.5$, the reverse occurs; the Stokes $v$ profile broadens, with the peak moving further from line center. Less circularly polarized flux is present in the bin $k$ from line center, and $R_0/R$ once again decreases.

\section{Conclusion}
\label{s:conc}

We have derived expressions for velocity subgroup populations in a one-dimensional maser saturated by either 1 or 2 beams of polarized radiation, described by the Stokes parameters. This theory has been coded in a new computer program, PRISM. Using this program,
we have demonstrated the expected amplification of circularly polarized Zeeman doublets and linearly polarized triplets in the large splitting case ($g \Omega / W > 1$), and that our code can show a smooth transition from this case to that of small splitting. In the small-splitting case, we show the appearance of the Van~Vleck angle at low amplification, the independence of the linear polarization fraction from the magnetic field strength, under moderate to strong saturation, and the generation of circular polarization at off-center frequencies. PRISM can consider non-zero Faraday rotation, and we demonstrate the rotation of the EVPA and suppression of linear polarization with increasing magnetic field at a fixed free electron number density. We compare our PRISM results to the analytic predictions of GKK, to the numerical calculations from \citet{2001ApJ...558L..55W}, and to the more recent work of LV19. We find our results to be compatible with these other works, given the differences of approach and levels of approximation. We discuss the limitations of our model and code with regards to saturation, and we also discuss the development with saturation of overall levels of circular and linear polarization.

\begin{acknowledgments}
This material is based on work supported by the National Science Foundation Graduate Research Fellowship Program under grant no. DGE—1144245. This research is also part of the Blue Waters sustained-petascale computing project, which is supported by the National Science Foundation (awards OCI-0725070 and ACI-1238993) the State of Illinois, and as of December, 2019, the National Geospatial-Intelligence Agency. Blue Waters is a joint effort of the University of Illinois at Urbana-Champaign and its National Center for Supercomputing Applications. MDG would like to thank The National Astronomy Research Institute of Thailand
(NARIT) for financial support and, for the early work on this project, the
UK STFC under consolidated grant ST/P000649/1 to the Jodrell Bank Centre for Astrophysics.
\end{acknowledgments}

\vspace{5mm}

\software{
Astropy \citep{2013A&A...558A..33A, 2018AJ....156..123A},
NumPy \citep{2020Natur.585..357H},
SciPy \citep{2020NatMe..17..261V},
Matplotlib \citep{2007CSE.....9...90H,2021zndo...5194481C},
tol-colors \citep{tol-colors-code,tol-colors-technote}
}

\appendix

\section{Some Derivation Details}
\label{a:appendix1}

We make a classical reduction of 
eq.(\ref{eq:dmoff1}) - eq.(\ref{eq:diag1pi}) that
has the following consequences:
(1) Different Fourier components of the radiation field are uncorrelated at
any degree of maser saturation; (2) There are no pulsations of the inversion,
so it is restricted to the central Fourier component with $n=0$; (3) The
statistics of the radiation are gaussian, and (4) Real lorentzians behave
as Dirac $\delta$-functions. Point (2) above dictates that $m=n$ in
eq.(\ref{eq:dmoff1}) and eq.(\ref{eq:dmoffpi1}), reducing 
the sums to single terms. The
sums survive in eq.(\ref{eq:diag1sig}) and 
eq.(\ref{eq:diag1pi}), but $n=0$, allowing many terms to be
combined. In particular, all off-diagonal DM elements now appear at Fourier
component $m$, allowing pairs of complex conjugate terms to be expressed as
real parts. The off-diagonal elements are then eliminated from
eq.(\ref{eq:diag1sig}) and eq.(\ref{eq:diag1pi})
using various versions of eq.(\ref{eq:dmoff1}) and eq.(\ref{eq:dmoffpi1}) with
the sums collapsed as described above. The resulting equations contain pairs
of complex field amplitudes that can be grouped and eliminated in favour of
the Stokes parameters. The resulting equations,
\begin{align}
\Delta_0^\pm   = \frac{P^\pm \phi (v)}{\Gamma^\pm}
              &  - \frac{\pi}{2 \hbar^2 \Gamma^\pm c \epsilon_0}
                 \sum_{m=-\infty}^\infty \Re
                 \left\{
                   2 | \hat{d}^0 |^2 \Delta_0^0 \tilde{L}_m^{0*}
                      \left[
                         {\cal I}_m - {\cal Q}_m \cos 2\phi - {\cal U}_m \sin 2 \phi
                      \right] \sin^2 \theta
                 \right. \nonumber \\
               &  \left.
                + 2 | \hat{d}^\pm |^2 \Delta_0^\pm \tilde{L}_m^\pm
                      \left[
                       (1+\cos^2 \theta ) {\cal I}_m
                      +({\cal Q}_m \cos 2\phi + {\cal U}_m \sin 2\phi )\sin^2 \theta
                      \mp 2 {\cal V}_m \cos \theta
                      \right]
                 \right. \nonumber \\
               &  \left.
                +   | \hat{d}^\mp |^2 \Delta_0^\mp \tilde{L}_m^\mp
                      \left[
                      (1+\cos^2 \theta ) {\cal I}_m
                      +({\cal Q}_m \cos 2\phi + {\cal U}_m \sin 2\phi )\sin^2 \theta
                      \pm 2 {\cal V}_m \cos \theta
                      \right]
                 \right\}
\label{eq:sigma_stokes}
\end{align}
and
\begin{align}
\Delta_0^0 = \frac{P^0 \phi(v)}{\Gamma^0} 
         & - \frac{\pi}{2 \hbar^2 \Gamma^0 c \epsilon_0}
             \sum_{m=-\infty}^\infty \Re
             \left\{
               4 | \hat{d}^0 |^2 \Delta_0^0 \tilde{L}_m^0
                  \left[
                    {\cal I}_m - {\cal Q}_m \cos 2\phi - {\cal U}_m \sin 2 \phi
                  \right] \sin^2 \theta
             \right. \nonumber \\
            &\left.
              + | \hat{d}^+ |^2 \Delta_0^+ \tilde{L}_m^{+*}
                  \left[
                       (1+\cos^2 \theta ) {\cal I}_m
                      +({\cal Q}_m \cos 2\phi + {\cal U}_m \sin 2\phi )\sin^2 \theta
                      - 2 {\cal V}_m \cos \theta
                  \right]
             \right. \nonumber \\
            &\left.
              + | \hat{d}^- |^2 \Delta_0^- \tilde{L}_m^{-*}
                  \left[
                       (1+\cos^2 \theta ) {\cal I}_m
                      +({\cal Q}_m \cos 2\phi + {\cal U}_m \sin 2\phi )\sin^2 \theta
                      + 2 {\cal V}_m \cos \theta
                  \right]
             \right\}
\label{eq:pi_stokes}
\end{align}
express the saturation of velocity-subgroup inversions by the Stokes parameters
of a ray in a single sample of duration equal to the reciprocal of the width of
the Fourier channels. Our Stokes parameters have the definitions
\begin{align}
 {\cal I}_m & = (1/2) c \epsilon_0 ( \amp{E}_{R,m} \amp{E}_{R,m}^*
                                    +\amp{E}_{L,m} \amp{E}_{L,m}^*) \\
 {\cal Q}_m & = (1/2) c \epsilon_0 ( \amp{E}_{R,m} \amp{E}_{L,m}^*
                                    +\amp{E}_{L,m} \amp{E}_{R,m}^*) \\
 {\cal U}_m & = (1/2)c \epsilon_0 i( \amp{E}_{R,m} \amp{E}_{L,m}^*
                                    -\amp{E}_{L,m} \amp{E}_{R,m}^*) \\
 {\cal V}_m & = (1/2) c \epsilon_0 ( \amp{E}_{R,m} \amp{E}_{R,m}^*
                                    -\amp{E}_{L,m} \amp{E}_{L,m}^*).
\end{align}

In equations \ref{eq:sigma_stokes} and \ref{eq:pi_stokes}, the only complex
quantities that remain inside the large braces are the lorentzian functions that
have the general definition
\begin{equation}
\tilde{L}_m^{\pm,0} = \frac{1}
       {2\pi[\gamma^{\pm,0} -i (\varpi_m - \Delta \omega^{\pm.0} -\omega_0 v/c)]},
\label{eq:complor}
\end{equation}
where the superscript $\pm,0$ represent optional transitions for selection. 
The Zeeman shifts for the three transitions 
are $\Delta \omega^\pm = \mp \Delta \omega$ for the $\sigma^\pm$ transitions
where $\Delta \omega$ is the absolute Zeeman shift. The $\pi$ transition
has $\Delta \omega^0 = 0$. It is
straightforward to show that the real part of eq.(\ref{eq:complor}) is equal
to half the real lorentzian,
\begin{equation}
L_m^{\pm,0} = \frac{\gamma^{\pm,0}/\pi}
                   {(\gamma^{\pm,0})^2
    + (\varpi_m - \Delta \omega^{\pm,0} -\omega_0 v/c )^2},
\label{eq:real_lor}
\end{equation}
so that the real part operation ($\Re$) on the contents of the braces 
in eq.(\ref{eq:sigma_stokes}) and eq.(\ref{eq:pi_stokes}) may be
carried out by replacing all complex lorentzians with their real counterparts, and
changing the 8 to 16 in the denominator multiplying the sums over $m$.

The analysis proceeds by noting that the classical reduction allows 
eq.(\ref{eq:real_lor}) to be used as a 
representation of the Dirac $\delta$-function, and this can
be used to collapse the $m$-sums in eq.(\ref{eq:sigma_stokes}) and 
eq.(\ref{eq:pi_stokes}). The $\delta$-function selects the transition, and velocity, dependent
Fourier component centered on 
the local frequency $\varpi^{\pm,0} = \Delta \omega^{\pm,0} + \omega_0 v/c$.
We use the shorthand expression ${\cal I}_{k^{\pm,0}} = {\cal I}(\varpi^{\pm,0})$, 
and similarly for the rest of the Stokes vector,
to represent these frequencies as indices on the Stokes parameters. The
somewhat reduced forms of eq.(\ref{eq:sigma_stokes}) and 
eq.(\ref{eq:pi_stokes}) are
\begin{align}
\Delta_0^\pm   = \frac{P^\pm \phi (v)}{\Gamma^\pm}
              &  - \frac{\pi}{4 \hbar^2 \Gamma^\pm c \epsilon_0}
                 \left\{
                   2 | \hat{d}^0 |^2 \Delta_0^0
                      \left[
          {\cal I}_{k^0} - {\cal Q}_{k^0} \cos 2\phi - {\cal U}_{k^0} \sin 2 \phi
                      \right] \sin^2 \theta
                 \right. \nonumber \\
               &  \left.
                + 2 | \hat{d}^\pm |^2 \Delta_0^\pm
                      \left[
    (1+\cos^2 \theta ) {\cal I}_{k^\pm}
     +({\cal Q}_{k^\pm} \cos 2\phi + {\cal U}_{k^\pm} \sin 2\phi )\sin^2 \theta
        \mp 2 {\cal V}_{k^\pm} \cos \theta
                      \right]
                 \right. \nonumber \\
               &  \left.
                +   | \hat{d}^\mp |^2 \Delta_0^\mp
                      \left[
           (1+\cos^2 \theta ) {\cal I}_{k^\mp}
     +({\cal Q}_{k^\mp} \cos 2\phi + {\cal U}_{k^\mp} \sin 2\phi )\sin^2 \theta
         \pm 2 {\cal V}_{k^\mp} \cos \theta
                      \right]
                 \right\}
\label{eq:sigma_k}
\end{align}
and
\begin{align}
\Delta_0^0 = \frac{P^0 \phi(v)}{\Gamma^0} 
         & - \frac{\pi}{4 \hbar^2 \Gamma^0 c \epsilon_0}
             \left\{
               4 | \hat{d}^0 |^2 \Delta_0^0
                  \left[
        {\cal I}_{k^0} - {\cal Q}_{k^0} \cos 2\phi - {\cal U}_{k^0} \sin 2 \phi
                  \right] \sin^2 \theta
             \right. \nonumber \\
            &\left.
              + | \hat{d}^+ |^2 \Delta_0^+
                  \left[
                       (1+\cos^2 \theta ) {\cal I}_{k^+}
        +({\cal Q}_{k^+} \cos 2\phi + {\cal U}_{k^+} \sin 2\phi )\sin^2 \theta
                      - 2 {\cal V}_{k^+} \cos \theta
                  \right]
             \right. \nonumber \\
            &\left.
              + | \hat{d}^- |^2 \Delta_0^-
                  \left[
                 (1+\cos^2 \theta ) {\cal I}_{k^-}
          +({\cal Q}_{k^-} \cos 2\phi + {\cal U}_{k^-} \sin 2\phi )\sin^2 \theta
                      + 2 {\cal V}_{k^-} \cos \theta
                  \right]
             \right\}.
 \label{eq:pi_k}   
\end{align}

The $\Delta_0^{\pm,0}$ in eq.(\ref{eq:sigma_k}) and eq.(\ref{eq:pi_k}) are
inversions in the velocity subgroup at velocity $v$. It is advantageous to
convert these into inversions, or molecular responses, at a particular frequency.
To this end, we multiply eq.(\ref{eq:sigma_k}) 
and eq.(\ref{eq:pi_k}) by the lorentzian-style $\delta$
function, 
\begin{equation}
L^{\pm,0}(v)\sim (\omega_0 /c)\delta (v\omega_0/c - [\varpi_n-\Delta \omega^{\pm,0}])
\label{eq:vlor}
\end{equation}
and integrate over all velocities. The result is to select a particular
velocity corresponding to the Zeeman-shifted frequency of the transition. If we
define the response as
\begin{equation}
\rho_0^{\pm,0}(\varpi) = \int_{-\infty}^\infty \Delta_0^{\pm,0}(v) L^{\pm,0}(v) dv
\label{eq:respodef}
\end{equation}
and note that the Zeeman shifts for each transition type are
$\Delta \omega^{\pm,0} = \mp \Delta \omega$ for the $\sigma$ transitions, where
$\Delta \omega$ is a positive definite frequency shift, and $\Delta \omega =0$ for
the $\pi$ transition, then we can re-cast eq.(\ref{eq:sigma_k}) 
and eq.(\ref{eq:pi_k}) respectively as
\begin{align}
&\rho_0^\pm  (\vpmz) = \frac{P^\pm \phi(\vpmz)}{\Gamma^\pm} \nonumber \\
                  & - \frac{\pi}{4 \hbar^2 \Gamma^\pm c \epsilon_0}
                 \left\{
                 2 | \hat{d}^0 |^2 \rho_0^0(\vpmz)
                      \left[
          {\cal I}(\vpmz) - {\cal Q}(\vpmz) \cos 2\phi - {\cal U}(\vpmz) \sin 2 \phi
                      \right] \sin^2 \theta
                 \right. \nonumber \\
               &  \left.
                +   | \hat{d}^\mp |^2 \rho_0^\mp(\vpmz)
                     \! \left[
           (1\!+\!\cos^2 \theta ) {\cal I}(\vpmt)
    \!+\!({\cal Q}(\vpmt)\cos 2\phi \!+ \!{\cal U}(\vpmt) \sin 2\phi )\sin^2 \theta
         \!\pm \!2 {\cal V}(\vpmt)\cos \theta
                      \right]
                 \right. \nonumber \\
               &  \left.
                + 2 | \hat{d}^\pm |^2 \rho_0^\pm(\vpmz)
                     \! \left[
    (1\!+\!\cos^2 \theta ) {\cal I}(\vpmz)
 \!+\!({\cal Q}(\vpn) \cos 2\phi \!+\! {\cal U}(\vpn) \sin 2\phi )\sin^2 \theta
        \!\mp\! 2 {\cal V}(\vpn) \cos \theta
                      \right]
            \!     \right\}
\label{eq:sigmadel}  
\end{align}
and
\begin{align}
&\rho_0^0  (\vpn) = \frac{P^0 \phi(\varpi)}{\Gamma^0} \nonumber \\
                  & - \frac{\pi}{4 \hbar^2 \Gamma^0 c \epsilon_0}
                 \left\{
        4 | \hat{d}^0 |^2 \rho_0^0(\vpn)
                  \left[
        {\cal I}(\vpn) - {\cal Q}(\vpn) \cos 2\phi - {\cal U}(\vpn) \sin 2 \phi
                  \right] \sin^2 \theta
             \right. \nonumber \\
            &\left.
              + | \hat{d}^- |^2 \rho_0^-(\vpn)
                  \left[
                 (1+\cos^2 \theta ) {\cal I}(\vpl)
    \!+\!({\cal Q}(\vpl) \cos 2\phi + {\cal U}(\vpl) \sin 2\phi )\sin^2 \theta
                      \!+\! 2 {\cal V}(\vpl) \cos \theta
                  \right]
             \right. \nonumber \\
            &\left.
              + | \hat{d}^+ |^2 \rho_0^+(\vpn)
                  \left[
                       (1+\cos^2 \theta ) {\cal I}(\vmi)
    \!+\!({\cal Q}(\vmi) \cos 2\phi + {\cal U}(\vmi) \sin 2\phi )\sin^2 \theta
                      \!-\! 2 {\cal V}(\vmi) \cos \theta
                  \right]
        \!    \right\}
\label{eq:pidel}
\end{align}
A formal realization average over the Stokes parameters, which can be isolated
statistically from the responses because of the assumption of gaussian statistics,
replaces the single realization versions with the averaged forms
$I(\vpmz)=\langle {\cal I}(\vpmz) \rangle$, and similarly for the other Stokes
parameters. Adoption of the shorthand notation $I(\vpmz)=I_{n\pm k}$, replacing
the Zeeman shift of $\Delta \omega$ by $k$ frequency bins returns the analysis
to eq.(\ref{eq:sig_key}) and eq.(\ref{eq:pi_key})
of the main text.

\section{Counter--Propagating Rays}
\label{a:bidirectional}

The solution presented in Section \ref{s:radtran} applies to a ray traveling in a given direction through the cloud, or Ray 1, in which the molecular populations have line of sight velocities denoted by $v$, corresponding to the angular frequency with respect to line center, $\varpi_n$. To apply this solution to a pair of counter--propagating rays, we must first consider how the solution changes for Ray 2, traveling in the opposite direction through the same population. Comparing Ray 2 to Ray 1, we must make the following adjustments:

\begin{enumerate}
    \item The angle of the magnetic field to the line of sight, $\theta_1$, is defined for Ray 1 as viewed from the end of the cloud that Ray 1 exits from. As $\theta$ preserves the directionality of the magnetic field, the corresponding angle between the magnetic field and line of sight for Ray 2 will be $\theta_2 = \pi + \theta_1$. Notably, the solution presented above only relies on $\theta$ in the form of $\sin^2 \theta$, $\cos^2 \theta$, and $\cos \theta$, so this point will only affect terms of un-squared $\cos \theta$, as $\cos \theta_2 = - \cos \theta_1$.
    \item The orientation of Stokes $u_1$ and $v_1$ as defined for Ray 1 correspond to the orientation of Stokes $-u_2$ and $-v_2$, respectively, for Ray 2. Stokes $i$ and $q$ are unchanged.
    \item Likewise, the generalized sky-plane angle, $\phi_1$, as defined for Ray 1 corresponds to the orientation of $-\phi_2$ as defined for Ray 2. The solution above only depends on $\phi$ in the form of $\cos (2 \phi )$ and $\sin (2 \phi) $. Therefore, this modification will only affect terms of $\sin(2 \phi)$ as $\sin(2 \phi_2) = - \sin( 2 \phi_1 )$.
\end{enumerate}

The Stokes parameters for each of the two rays are first calculated separately following Equation \ref{eq:fullformsol}. For Ray 1, the dimensionless gain matrix terms are given as defined in Equations \ref{eq:gi_dimless} - \ref{eq:qu_dimless}. Defining all parameters discussed above by their Ray 1 values, the first two dimensionless gain matrix terms for Ray 2, $\gamma_i$ and $\gamma_q$, do not change from their values for Ray 1. The remaining terms become:

\begin{eqnarray}
\gamma_{u,n,2} &=& - \left( 2 \delta_{n}^0  - \eta^+ \delta_{n+k}^+ - \eta^- \delta_{n-k}^-  \right) \sin^2 \theta \sin \left( 2 \phi \right) \label{eq:gu_dimless_ray2} \\
\gamma_{v,n,2} &=&   - 2 \left( \eta^+ \delta_{n+k}^+ - \eta^- \delta_{n-k}^- \right) \cos \theta \label{eq:gv_dimless_ray2} \\
\gamma_{qu,n,2} &=& + \dfrac{ 4 \Gamma w \sqrt{\pi} \nu_c e^3 n_e B \cos \theta}{3\sqrt{\pi} A^0 P^0 \epsilon_0 m_e^2 c^4 } \label{eq:qu_dimless_ray2}
\end{eqnarray}

In addition, when calculating the Stokes parameters for Ray 2, Equation \ref{eq:fullformsol} must be integrated in the opposite direction.

As the population at each $(\varpi_n,\tau)$ can only have one batch of effective Stokes parameters, we define the effective Stokes as the sum of the Stokes parameters from the two different rays. This is because the intensity at a given point $(\varpi_n,\tau)$ is determined by the buildup of light from each of the two directions. However, according to point 2 above, Stokes $u_2$ and $v_2$ have to be inverted to be combined with Stokes $u_1$ and $v_1$:
\begin{eqnarray}
    i_n (\tau) &=& i_{n,1} (\tau) + i_{n,2} (\tau) \label{eq:sum-stoki}\\
    q_n (\tau) &=& q_{n,1} (\tau) + q_{n,2} (\tau)  \label{eq:sum-stokq}\\
    u_n (\tau) &=& u_{n,1} (\tau) - u_{n,2} (\tau)  \label{eq:sum-stoku}\\
    v_n (\tau) &=&  v_{n,1} (\tau) - v_{n,2} (\tau)  \label{eq:sum-stokv}
\end{eqnarray}
as defined in the reference frame of Ray 1. Of course, the Ray 2 Stokes parameters are integrated in tau starting from the opposite end of the maser as the Ray 1 Stokes parameters.

Since the unitless inversions represent the molecular energy state of the population, there must also be one unified value for each inversion at any point in the $(\varpi_n,\tau)$ parameter space. These $\delta^{\pm,0}_n (\tau)$ are the values that are being solved for within the system of equations. While the formulae that are used to calculate them (Equations \ref{eq:sig_dimless} and \ref{eq:pi_dimless}) utilize the total effective stokes parameters from equations (\ref{eq:sum-stoki}) - (\ref{eq:sum-stokv}), they are still dependent on the angles $\theta$ and $\phi$, which are different for each of the two rays. 

However, a closer inspection of Equations \ref{eq:sig_dimless} and \ref{eq:pi_dimless} reveals that the only resulting terms that change signs between Rays 1 and 2, $\cos \theta$ and $\sin(2\phi)$, only appear when multiplied by Stokes $u$ and $v$, respectively, which also change signs between reference frames. Therefore, the effective inversion as calculated from the effective Stokes parameters for both rays combined, $\delta_n^{\pm,0}(\tau)$, is the same regardless of the reference frame in which it was calculated.

\section{Loss of Van~Vleck Angle at low intensity}
\label{appendix3}

The presence of a Van~Vleck angle, at which $m_l$ is a minimum, between field angles of $0$ and $\pi/2$ is a feature of a wide range of conditions encompassing low to moderate saturation that are discussed in more detail in Section~\ref{ss:siopolnof}. The establishment of the angle for a constant magnitude magnetic field and increasing optical depth is shown graphically in Figure~\ref{fig:mlevpa_v_theta_nofr_k01}. We note that for very strong saturation, LV19 expect the Van~Vleck angle to disappear again. We prove here that there is no Van~Vleck angle in the limit of low intensity, in the sense that the gain coefficients in all the Zeeman transitions remain approximately constant.

In the case where there is no Faraday rotation, $\sin(2\phi)=0$, and zero Stokes-$V$ at line centre, the following equations for propagation in the central frequency bin ($n=0$) can be derived from eq.(\ref{eq:rteq}) and the gain matrix in eq.(\ref{eq:gainmat}) with components from eq.(\ref{eq:gammai}) and eq.(\ref{eq:gammaq}):
\begin{equation}
di/ds = [(1+\cos^2 \theta) i + q \sin^2 \theta](\gamma_k^+ + \gamma_{-k}^{-})
      + 2\gamma_0 (i-q)\sin^2 \theta
\end{equation}    
\begin{equation} 
dq/ds = [(1+\cos^2 \theta) q + i \sin^2 \theta](\gamma_k^+ + \gamma_{-k}^{-})
       + 2 \gamma_0 (q-i) \sin^2 \theta ,
\end{equation}
where the $n=0$ subscript has been dropped for brevity. The gain coefficients in the equations above follow the pattern 
$\gamma_k^+ = \pi |\hat{d}^+|^2 \rho_k^+ /(4 \epsilon_0 \hbar)$. The two propagation equations above can be added and subtracted to form equations in the sum and difference of the Stokes parameters. With the assumption that the gain coefficients are constant (very weak saturation) the sum and difference equations may be integrated to yield,
\begin{equation}
  (i+q)(s) = (i+q)_0\exp[2 \beth s] , 
  \label{eq:app_prop1}
\end{equation}
where $\beth = \gamma_k^+ + \gamma_{-k}^-$, and
\begin{equation}
 (i-q)(s) = (i-q)_0 \exp[2(2\gamma_0 \sin^2 \theta + \beth \cos^2 \theta)s].
 \label{eq:app_prop2}
\end{equation}
Subtraction of eq.(\ref{eq:app_prop2}) from eq.(\ref{eq:app_prop1}), and the
assumption of an unpolarized background of intensity $i_0 = i_{BG}$, leads to the following equation in $q(s)$ only:
\begin{equation}
q(s) = (i_{BG}/2) e^{2\beth s} \left[
  1 - e^{4 \gamma_0 s \sin^2 \theta} e^{2 \beth s (\cos^2 \theta - 1)}
                               \right].
\label{eq:app_qsoln}
\end{equation}
Equation~\ref{eq:app_qsoln} may be differentiated with respect to $\theta$, and the
result equated to zero to investigate turning points. These are found only at 
angles satisfying $\sin(2\theta)=0$, so there are no turning points in $q$ between
$\theta =0$ and $\theta = \pi/2$ except at the end-points of the range. It is also straightforward to show that there are no zeros of $q$ in the same range by setting the left-hand side of eq.(\ref{eq:app_qsoln}) to zero and then taking the logarithm of the expression in square brackets. In this case a common factor of $\sin^2 \theta$ can be extracted, so there are also no zeros in the selected range of the angle.

\bibliography{everything,software}
\bibliographystyle{aasjournal}

\end{document}